\documentclass[twocolumn]{aastex631}

\usepackage{amsmath}
\usepackage{amssymb}
\usepackage{graphicx,float}
\usepackage[utf8]{inputenc}
\usepackage{booktabs}
\usepackage{array,multirow}
\usepackage{gensymb}
\usepackage{float}

\usepackage{CJKutf8}

\begin{document}
\begin{CJK*}{UTF8}{gbsn}
\title{\large A Search for Millimeter-Bright Blazars as Astrophysical Neutrino Sources }

\affiliation{III. Physikalisches Institut, RWTH Aachen University, D-52056 Aachen, Germany}
\affiliation{Department of Physics, University of Adelaide, Adelaide, 5005, Australia}
\affiliation{Dept. of Physics and Astronomy, University of Alaska Anchorage, 3211 Providence Dr., Anchorage, AK 99508, USA}
\affiliation{School of Physics and Center for Relativistic Astrophysics, Georgia Institute of Technology, Atlanta, GA 30332, USA}
\affiliation{Dept. of Physics, Southern University, Baton Rouge, LA 70813, USA}
\affiliation{Dept. of Physics, University of California, Berkeley, CA 94720, USA}
\affiliation{Lawrence Berkeley National Laboratory, Berkeley, CA 94720, USA}
\affiliation{Institut f{\"u}r Physik, Humboldt-Universit{\"a}t zu Berlin, D-12489 Berlin, Germany}
\affiliation{Fakult{\"a}t f{\"u}r Physik {\&} Astronomie, Ruhr-Universit{\"a}t Bochum, D-44780 Bochum, Germany}
\affiliation{Universit{\'e} Libre de Bruxelles, Science Faculty CP230, B-1050 Brussels, Belgium}
\affiliation{Vrije Universiteit Brussel (VUB), Dienst ELEM, B-1050 Brussels, Belgium}
\affiliation{Dept. of Physics, Simon Fraser University, Burnaby, BC V5A 1S6, Canada}
\affiliation{Department of Physics and Laboratory for Particle Physics and Cosmology, Harvard University, Cambridge, MA 02138, USA}
\affiliation{Dept. of Physics, Massachusetts Institute of Technology, Cambridge, MA 02139, USA}
\affiliation{Dept. of Physics and The International Center for Hadron Astrophysics, Chiba University, Chiba 263-8522, Japan}
\affiliation{Department of Physics, Loyola University Chicago, Chicago, IL 60660, USA}
\affiliation{Dept. of Physics and Astronomy, University of Canterbury, Private Bag 4800, Christchurch, New Zealand}
\affiliation{Dept. of Physics, University of Maryland, College Park, MD 20742, USA}
\affiliation{Dept. of Astronomy, Ohio State University, Columbus, OH 43210, USA}
\affiliation{Dept. of Physics and Center for Cosmology and Astro-Particle Physics, Ohio State University, Columbus, OH 43210, USA}
\affiliation{Niels Bohr Institute, University of Copenhagen, DK-2100 Copenhagen, Denmark}
\affiliation{Dept. of Physics, TU Dortmund University, D-44221 Dortmund, Germany}
\affiliation{Dept. of Physics and Astronomy, Michigan State University, East Lansing, MI 48824, USA}
\affiliation{Dept. of Physics, University of Alberta, Edmonton, Alberta, T6G 2E1, Canada}
\affiliation{Erlangen Centre for Astroparticle Physics, Friedrich-Alexander-Universit{\"a}t Erlangen-N{\"u}rnberg, D-91058 Erlangen, Germany}
\affiliation{Physik-department, Technische Universit{\"a}t M{\"u}nchen, D-85748 Garching, Germany}
\affiliation{D{\'e}partement de physique nucl{\'e}aire et corpusculaire, Universit{\'e} de Gen{\`e}ve, CH-1211 Gen{\`e}ve, Switzerland}
\affiliation{Dept. of Physics and Astronomy, University of Gent, B-9000 Gent, Belgium}
\affiliation{Dept. of Physics and Astronomy, University of California, Irvine, CA 92697, USA}
\affiliation{Karlsruhe Institute of Technology, Institute for Astroparticle Physics, D-76021 Karlsruhe, Germany}
\affiliation{Karlsruhe Institute of Technology, Institute of Experimental Particle Physics, D-76021 Karlsruhe, Germany}
\affiliation{Dept. of Physics, Engineering Physics, and Astronomy, Queen's University, Kingston, ON K7L 3N6, Canada}
\affiliation{Department of Physics {\&} Astronomy, University of Nevada, Las Vegas, NV 89154, USA}
\affiliation{Nevada Center for Astrophysics, University of Nevada, Las Vegas, NV 89154, USA}
\affiliation{Dept. of Physics and Astronomy, University of Kansas, Lawrence, KS 66045, USA}
\affiliation{Centre for Cosmology, Particle Physics and Phenomenology - CP3, Universit{\'e} catholique de Louvain, Louvain-la-Neuve, Belgium}
\affiliation{Department of Physics, Mercer University, Macon, GA 31207-0001, USA}
\affiliation{Dept. of Astronomy, University of Wisconsin{\textemdash}Madison, Madison, WI 53706, USA}
\affiliation{Dept. of Physics and Wisconsin IceCube Particle Astrophysics Center, University of Wisconsin{\textemdash}Madison, Madison, WI 53706, USA}
\affiliation{Institute of Physics, University of Mainz, Staudinger Weg 7, D-55099 Mainz, Germany}
\affiliation{Department of Physics, Marquette University, Milwaukee, WI 53201, USA}
\affiliation{Institut f{\"u}r Kernphysik, Universit{\"a}t M{\"u}nster, D-48149 M{\"u}nster, Germany}
\affiliation{Bartol Research Institute and Dept. of Physics and Astronomy, University of Delaware, Newark, DE 19716, USA}
\affiliation{Dept. of Physics, Yale University, New Haven, CT 06520, USA}
\affiliation{Columbia Astrophysics and Nevis Laboratories, Columbia University, New York, NY 10027, USA}
\affiliation{Dept. of Physics, University of Oxford, Parks Road, Oxford OX1 3PU, United Kingdom}
\affiliation{Dipartimento di Fisica e Astronomia Galileo Galilei, Universit{\`a} Degli Studi di Padova, I-35122 Padova PD, Italy}
\affiliation{Dept. of Physics, Drexel University, 3141 Chestnut Street, Philadelphia, PA 19104, USA}
\affiliation{Physics Department, South Dakota School of Mines and Technology, Rapid City, SD 57701, USA}
\affiliation{Dept. of Physics, University of Wisconsin, River Falls, WI 54022, USA}
\affiliation{Dept. of Physics and Astronomy, University of Rochester, Rochester, NY 14627, USA}
\affiliation{Department of Physics and Astronomy, University of Utah, Salt Lake City, UT 84112, USA}
\affiliation{Dept. of Physics, Chung-Ang University, Seoul 06974, Republic of Korea}
\affiliation{Oskar Klein Centre and Dept. of Physics, Stockholm University, SE-10691 Stockholm, Sweden}
\affiliation{Dept. of Physics and Astronomy, Stony Brook University, Stony Brook, NY 11794-3800, USA}
\affiliation{Dept. of Physics, Sungkyunkwan University, Suwon 16419, Republic of Korea}
\affiliation{Institute of Physics, Academia Sinica, Taipei, 11529, Taiwan}
\affiliation{Dept. of Physics and Astronomy, University of Alabama, Tuscaloosa, AL 35487, USA}
\affiliation{Dept. of Astronomy and Astrophysics, Pennsylvania State University, University Park, PA 16802, USA}
\affiliation{Dept. of Physics, Pennsylvania State University, University Park, PA 16802, USA}
\affiliation{Dept. of Physics and Astronomy, Uppsala University, Box 516, SE-75120 Uppsala, Sweden}
\affiliation{Dept. of Physics, University of Wuppertal, D-42119 Wuppertal, Germany}
\affiliation{Deutsches Elektronen-Synchrotron DESY, Platanenallee 6, D-15738 Zeuthen, Germany}

\author[0000-0001-6141-4205]{R. Abbasi}
\affiliation{Department of Physics, Loyola University Chicago, Chicago, IL 60660, USA}

\author[0000-0001-8952-588X]{M. Ackermann}
\affiliation{Deutsches Elektronen-Synchrotron DESY, Platanenallee 6, D-15738 Zeuthen, Germany}

\author{J. Adams}
\affiliation{Dept. of Physics and Astronomy, University of Canterbury, Private Bag 4800, Christchurch, New Zealand}

\author[0000-0002-9714-8866]{S. K. Agarwalla}
\altaffiliation{also at Institute of Physics, Sachivalaya Marg, Sainik School Post, Bhubaneswar 751005, India}
\affiliation{Dept. of Physics and Wisconsin IceCube Particle Astrophysics Center, University of Wisconsin{\textemdash}Madison, Madison, WI 53706, USA}

\author[0000-0003-2252-9514]{J. A. Aguilar}
\affiliation{Universit{\'e} Libre de Bruxelles, Science Faculty CP230, B-1050 Brussels, Belgium}

\author[0000-0003-0709-5631]{M. Ahlers}
\affiliation{Niels Bohr Institute, University of Copenhagen, DK-2100 Copenhagen, Denmark}

\author[0000-0002-9534-9189]{J.M. Alameddine}
\affiliation{Dept. of Physics, TU Dortmund University, D-44221 Dortmund, Germany}

\author{N. M. Amin}
\affiliation{Bartol Research Institute and Dept. of Physics and Astronomy, University of Delaware, Newark, DE 19716, USA}

\author[0000-0001-9394-0007]{K. Andeen}
\affiliation{Department of Physics, Marquette University, Milwaukee, WI 53201, USA}

\author[0000-0003-4186-4182]{C. Arg{\"u}elles}
\affiliation{Department of Physics and Laboratory for Particle Physics and Cosmology, Harvard University, Cambridge, MA 02138, USA}

% ...

\collaboration{1000}{The IceCube Collaboration}

\author[0000-0003-1690-6678]{A.~D.~Hincks}\affiliation{David A. Dunlap Department of Astronomy and Astrophysics, University of Toronto, 50 St. George St., Toronto ON M5S 3H4, Canada}\affiliation{Specola Vaticana (Vatican Observatory), V-00120 Vatican City State}

\author[0000-0003-2622-6895]{X.~Ma (马潇依)}\affiliation{Kavli Institute for Astronomy and Astrophysics, Peking University, Beijing 100871, China}\affiliation{Department of Astronomy, School of Physics, Peking University, Beijing 100871, China}

\author[0000-0001-5327-1400]{C.~Vargas}\affiliation{Department of Physics \& Astronomy, Texas A\&M University, College Station, TX 77843, USA}\affiliation{Mitchell Institute for Fundamental Physics \& Astronomy, Texas A\&M University, College Station, TX 77843, USA}

\author[0000-0002-4765-3426]{C.~Herv\'ias-Caimapo}\affiliation{Instituto de Astrof\'isica and Centro de Astro-Ingenier\'ia, Facultad de F\'isica, Pontificia Universidad Cat\'olica de Chile, Chile}

\author[0000-0001-5210-7625]{E.~S.~Battistelli}\affiliation{Sapienza Universit\`a di Roma, Piazzale Aldo Moro, 5, 00185, Rome (RM), Italy}

\author[0000-0001-7109-0099]{K.~Huffenberger}\affiliation{Department of Physics \& Astronomy, Texas A\&M University, College Station, TX 77843, USA}\affiliation{Mitchell Institute for Fundamental Physics \& Astronomy, Texas A\&M University, College Station, TX 77843, USA}\affiliation{Department of Physics, Florida State University, Tallahassee, FL 32306 USA}

\author[0000-0002-4478-7111]{S.~Naess}\affiliation{Institute for theoretical astrophysics, University of Oslo, Norway}

\author[0000-0003-1842-8104]{J.~Orlowski-Scherer}\affiliation{Department of Physics and Astronomy, University of Pennsylvania, Philadelphia, PA 19104 USA}

\author[0000-0001-6541-9265]{B.~Partridge}\affiliation{Department of Physics and Astronomy, Haverford College, 370 Lancaster Ave, Haverford, PA 19041, USA}

\author[0000-0002-8149-1352]{Crist\'obal Sif\'on}\affiliation{Instituto de F\'isica, Pontificia Universidad Cat\'olica de Valpara\'iso, Casilla 4059, Valpara\'iso, Chile}

\author[0000-0002-7567-4451]{E.~J.~Wollack}\affiliation{NASA/Goddard Space Flight Center, Greenbelt, MD, USA 20771}

\collaboration{1000}{The Atacama Cosmology Telescope Collaboration}

\date{\today}

\begin{abstract}
The powerful jets of blazars have been historically considered as likely sites of high-energy cosmic-ray acceleration. However, particulars of the launched jet and the locations of leptonic and hadronic jet loading remain unclear. In the case when leptonic and hadronic particle injection occur jointly, a temporal correlation between synchrotron radiation and neutrino production is expected. We use a first catalog of millimeter (mm) wavelength blazar light curves from the Atacama Cosmology Telescope for a time-dependent correlation with twelve years of muon neutrino events from the IceCube South Pole Neutrino Observatory. Such mm emission is known to trace activity of the bright jet base, which is often self-absorbed at lower frequencies and potentially gamma-ray opaque. We perform an analysis of the population, as well as analyses of individual, selected sources. We do not observe a significant signal from the stacked population. TXS 0506+056 is found as the most significant, individual source, though this detection is not globally significant in our analysis of selected AGN.  Our results suggest that the majority of mm-bright blazars are neutrino dim. In general, it is possible that many blazars have lighter, leptonic jets, or that only selected blazars provide exceptional conditions for neutrino production.  
\end{abstract}

\section{Introduction}
\subsection{Blazars as High-Energy Accelerators}
The relativistic jets of active galactic nuclei (AGN), the matter-accreting supermassive black holes at the centers of galaxies, may act as sites of hadronic cosmic-ray acceleration \citep{hovatta2020relativisticjetsblazars, Murase_2012}. Blazar AGN are characterized by jets oriented within a few degrees of the line of sight to Earth. Their Doppler-boosted emission makes them especially bright high-energy sources, in spite of their distance. However, the dynamics and energy budgets of these systems are not well constrained \citep{hovatta2020relativisticjetsblazars}. Additionally, the charged nuclei potentially accelerated at these sources are deflected in magnetic fields along the line of sight, degrading their directional information \citep{Allard_2012}. Alternatively, high-energy neutrinos may be produced in hadronic and photohadronic collisions along the jet. Progress can be made by combining observations of the radiative emission from the jet with measurements of high-energy neutrinos to understand the relationship between nonthermal processes and hadronic particle interactions. This complete multimessenger picture may ultimately allow us to infer the underlying hadronic populations, their acceleration, transport and jet feeding within blazars.

The IceCube South Pole Neutrino Observatory began data collection with a completed detector array in December of 2010. In 2013, an isotropic excess of high-energy events was observed over the expected lower-energy atmospheric cosmic-ray background. This was identified as an all-sky flux of astrophysical neutrinos \citep{Aartsen_2013a}. Only in recent years have sufficient neutrinos accumulated that a significant signal could be identified from a small number of the brightest, individual sources over background. While several contributing sources have been established -- emerging Seyfert galaxies \citep{ngc_1068, x_ray_bright_seyferts}, emission from the Galactic Plane \citep{galactic_plane}, and selected blazar AGN \citep{txs_time_int, Aartsen_2020, ngc_1068} -- the exact composition of the astrophysical diffuse flux is unknown. 

The first candidate neutrino source, TXS~0506+056, is a radio-bright blazar AGN \citep{Kun_2018}. This is the central supermassive black hole of a galaxy at redshift 0.337 \citep{txs_redshift}. As is characteristic of blazars, a beamed jet is observed, aligned within several degrees of the line of sight to Earth. It was first identified as a likely neutrino source in 2017 when a $\sim$300 TeV event was observed from the source with low angular uncertainty, coincident with a six-month gamma-ray flare \citep{Fermi_TXS_ATel, 2018Sci...361.1378I}.  After this multimessenger discovery in 2017, an archival search through existing IceCube data was performed, revealing a significant five-month neutrino flare beginning in 2014 \citep{txs_time_int}. However, this earlier period is not known to have been accompanied by an increase in either gamma-ray or radio emission. Since then, a handful of additional blazars have begun to emerge as individual point sources with approximate, local $\sim$$3\sigma$ significance, including PKS~1424+240 and GB6~J1542+6129 \citep{Aartsen_2020}. These significances reflect only the trial of the specific source location, and not the larger trials correction associated with testing the full source catalog. 

A number of analyses have searched for emission from the population of observed blazars. Recent correlations with the Fermi-LAT 3FHL and 1FLE catalogs have targeted gamma-ray-bright blazars and have generally set limits with an integral neutrino flux less than a tenth of the total astrophysical diffuse emission observed by IceCube \citep{search_3fhl, search_1fle}. Studies have also focused on radio-bright blazars, some suggestive of a potential correlation with the core or jet base \citep{plavin2020, radio_2021, owens_metsahovi, antares1, antares2, Plavin_2023}. A recent IceCube analysis of 15\,GHz Monitoring of Jets in Active Galactic Nuclei with VLBA Experiments (MOJAVE) AGN found a small excess from the source population at the level of 1.9$\sigma$ \citep{mojave}. This result was consistent with a contribution to less than ten percent of the diffuse astrophysical flux after correcting for the completeness of the catalog. In this work, we investigate a potential correlation with shorter-wavelength millimeter (mm) blazar activity using data from the Atacama Cosmology Telescope (ACT), a cosmic microwave background (CMB) experiment that surveyed ${\sim}45\%$ of the sky over several years at mid- and southern latitudes.

\subsection{Leptonic and Hadronic Loading of the Jet}

Multiple mechanisms have been proposed for jet production from compact, astrophysical objects. In the case of blazars, these high powers are often attributed to the Blandford-Znajek process \citep{Blandford:1977ds}. The ionized current flow of the accretion disk would establish a poloidal field around the black hole. Disturbances from the external environment may lead to disorganization of the field structure and stray lines. As the black hole rotates, the winding of these magnetic fields leads to helical or toroidal structures, supporting the outflow of material from the inner parsecs as a jet stream. 

The `core' of the blazar commonly refers to the radio-opaque region near the base of the jet. Electrons in the region of the accretion disk may be accelerated in shocks or magnetic reconnection events \citep{Matthews_2020}. These leptonic populations are loaded into the jet, producing synchrotron radiation. To remain in thermal equilibrium, photons in these regions of higher density may be reabsorbed onto the parent electron population, a process called ``self-absorption". The longest radio wavelengths of synchrotron spectra typically become self-absorbed first, forming an opaque region where the conditions for free streaming are only achieved at the edge \citep{early_jet_survey, jets_as_radio_sources}. Shorter-wavelength mm emission is less absorbed, providing a preferable tracer of activity from sub-parsec-scale regions forming the base of the jet. 

At higher energies, radiation from both accelerated electron and proton populations can dominate the observed SED. Inverse Compton scattering should contribute in denser, kinetically dominated regions, though X-rays and gamma-rays may also be produced by hadronic processes \citep{Cerruti_2020, Petropoulou_2014}. We note that abundant photon fields could create a gamma-ray opaque environment, where the collision of photons leads to pair production, cascading the observed emission to lower MeV-energies. 

Astrophysical neutrino production requires hadronic processes and is primarily anticipated through two channels. In dense regions of accelerated protons, proton--proton ($p$--$p$) collisions may efficiently yield neutrinos through pion production and decay. Within the bases of blazar jets, $p$--$\gamma$ collisions may be more common, since this part of the jet acts as a natural accelerator of protons, and both external fields of photons and those produced within the jet are present:

\begin{equation}
    p + \gamma \rightarrow \Delta^{+} \rightarrow 
    \begin{cases}
      \pi^{+} + n & \text{1/3 B.R.}\\
      \pi^{0} + p & \text{2/3 B.R.}\\
    \end{cases}       
\end{equation}
Notably, this production channel only becomes efficient as the center of momentum energy of the collision exceeds the $\Delta$(1232) resonance energy of 1.23 GeV. Details of these production scenarios can be found in existing literature \citep[e.g.,][]{Murase_2023}. In either case, after oscillating over the long distances between astrophysical sources and Earth, the expected rates are essentially equal between the three neutrino flavors. 

The environment and loaded material at the base of the jet are not well understood. If electrons and protons are loaded into the jet together, radiation from leptonic processes might facilitate or trace neutrino production. Still, the distance of the jet base from the black hole, its size, opening angle, magnetic field structure, external photon fields and bulk Lorentz factor are not easy to establish. As the injected proton and electron spectra are also poorly constrained, models of the expected neutrino emission from blazar jet flares carry large systematic uncertainties. 

A correlation between gamma-ray and radio flares has been established \citep{Fuhrmann_2014}. The strength of the correlation was seen to increase with radio frequency, suggesting that mm emission from the core may be a good tracer of high-energy particle production. In the case of a gamma-opaque production zone, unabsorbed mm emission would continue to provide this indicator of particle loading. 

In this work, a model is considered in which the injection of electrons and protons leads to temporally-correlated synchrotron and neutrino flares. A proportionality is assumed between mm flux density and neutrino intensity. This scenario requires that the Compton-dominance of the emissive region at the base of the jet allows for bright, observable radio/mm flares \citep{Boula_2018, Wang_2022}. Additionally, sufficiently accelerated protons must be present to interact with either radiation from within the jet or from external fields. In this way, synchrotron radiation may act as either a tracer or facilitator of neutrino production. While the relationship between mm flux and neutrino intensity is likely a function of time, depending on the evolution of the flaring region and individual jet environment, we adopt this simple, fixed proportionality as a first approximation.

Our paper is organized as follows. We first present the ACT catalog of blazars in Sec.~\ref{sec:act_data}, where we include a discussion of TXS~0506+056 (Sec.~\ref{ssec:txs}), a likely blazar neutrino source. In this section, we also describe the temporal models constructed with the ACT light curves (Sec.~\ref{ssec:neutrino_mm_corr}). We describe IceCube data acquisition and data products in Sec.~\ref{sec:icecube_data}. We then proceed to describe our methodology for searching for a mm--neutrino correlation and report the results in Sec.~\ref{sec:method_results}, both with individual blazars (Sec.~\ref{ssec:individual}) and with the population as a whole (Sec.~\ref{ssec:population}). We discuss our results and conclude in Sec.~\ref{sec:discussion}.

\section{A First Catalog of ACT Millimeter-Wavelength Blazar Light Curves}\label{sec:act_data}

The Atacama Cosmology Telescope (ACT) was a six meter telescope in the Atacama Desert of northern Chile built to image the CMB with $\sim$arcminute resolution.  Over its 15 year lifetime, it observed a number of fields spanning declinations from $-62\degree$ to $+22\degree$. The first two generations of receiver---the Millimeter Bolometric Array Camera (MBAC, 2008--2010; \citealt{Swetz_2011}) and ACTPol (2013--2016; \citealt{thornton/etal:2016})---observed eight fields ranging from hundreds to thousands of square degrees. In 2016, a year before the third generation receiver, Advanced ACTPol (AdvACT, 2017--22; \citealt{henderson/etal:2016}), was installed, a wide, $\sim$18,000\,deg$^2$ survey was undertaken, covering the full declination range and a wide range of right ascensions that avoided the Galactic plane \citep{naess/etal:2020}. The survey lasted until the observatory was decommissioned in September 2022 and was carried out in three principal frequency bands: 77--112\,GHz (95.0\,GHz), 124--172\,GHz (146.9\,GHz) and 182--277\,GHz (225.0\,GHz), where the values in parentheses are band centers appropriate for synchrotron radiation with spectral index, $\alpha = 0.7$. Throughout, we refer to these bands by their central values of 95, 147 and 225\,GHz. Two low frequency bands, 21--32\,GHz and 29--48\,GHz, were added in 2020; these centimeter wavelength observations are still being characterized and are not included in this paper.

\begin{figure*}
\centering
\includegraphics[width=\linewidth]{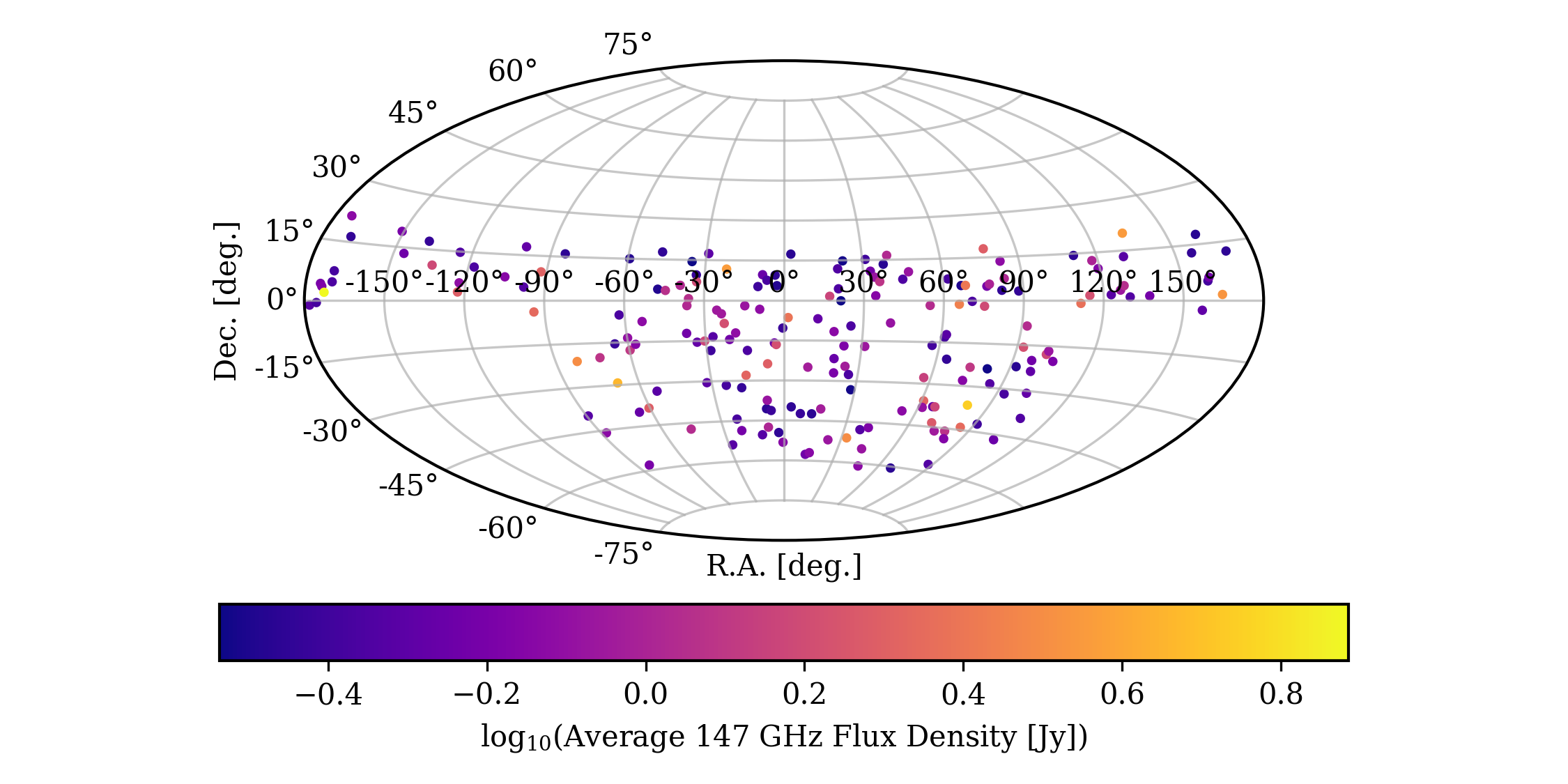}
\caption{Map showing the spatial distribution in equatorial coordinates of the 195 blazars in the ACT catalog of mm-bright AGN. The average 147\,GHz flux over the observing period is indicated by the marker color. ACT surveyed from $+22^{\circ}$ in the north to $-62\degree$ in the southern hemisphere.}
\label{fig:act_catalog_map}
\end{figure*}

The ACT Collaboration has produced a first catalog of 95, 147 and 225\,GHz light curves for the brightest mm point sources in the ACTPol and AdvACT surveys. A paper describing the catalog in detail is in preparation (Ma et al.) and the light curves will be publicly released. Here, we summarize some of the basic properties of the data. The catalog consists of 205 radio sources with mean fluxes above 500\,mJy in the 95\,GHz band. This flux density threshold results in high signal-to-noise light curves. Individual flux-density measurements are generally associated with few-percent statistical and systematic uncertainties. Future processing and data calibration performed by ACT may shift the flux-density amplitude of the final light curve products within this range. Some sources are observed from 2013 onwards, but data for most sources date from 2016 when the wide-area survey began. The beam size is approximately 1.4 arcminutes at 147 GHz and the point-source positional uncertainty is on the order of 3 arcseconds \citep{gralla}. All sources have been cross-matched in the VizieR database, the NVSS databases, or the AT20G Australia Telescope 20\,GHz survey catalog, and 195 of them are confirmed blazars \citep{vizier, nvss, at20g}. The positions from these external catalogs are adopted throughout this work because of their much smaller uncertainties. Fig.~\ref{fig:act_catalog_map} shows a map of source position and average mm flux density.

\begin{figure*}
\centering
\includegraphics[width=\linewidth]{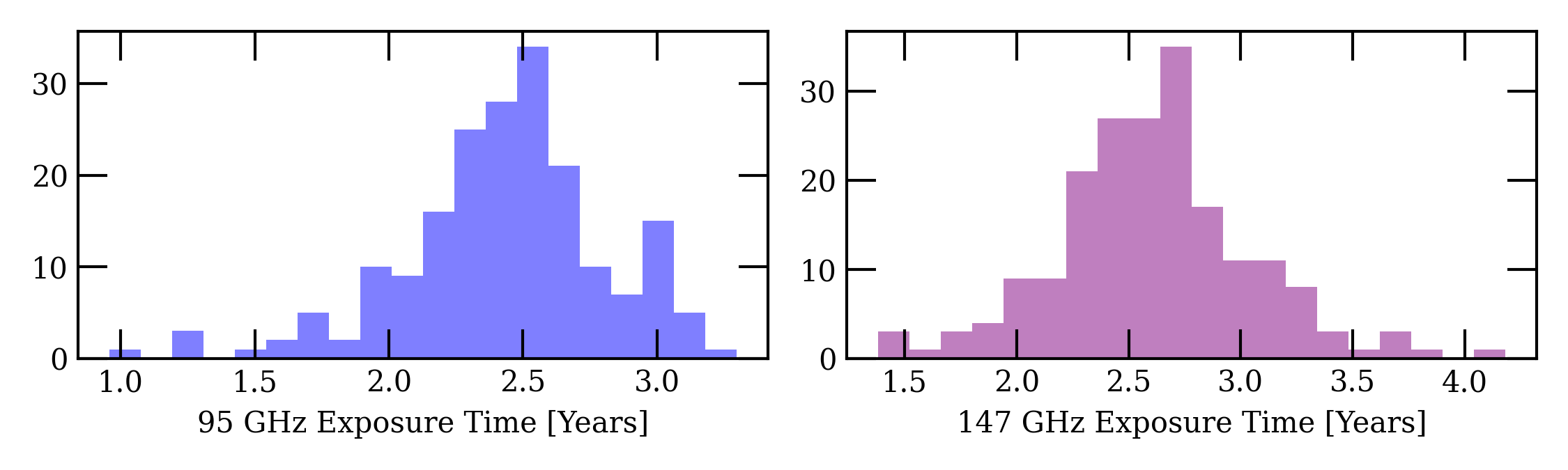}
\caption{Histograms showing the amount of time each of the 195 blazars in the sample was observed, for the 95 and 147\,GHz bands. Each data point in a source's light curve is assumed to represent its flux for fourteen days surrounding the date it was made on. The union of this set is expressed as its exposure time. We do not expect substantial mm jet variability at this or shorter time scales. ACT takes few-month pauses for seasonal weather conditions or other telescope maintenance. We also note that there is some spread in the total exposure time of each source. Still, these linearly-interpolated light curves provide a useful characterization of source activity. The majority of sources at 147 GHz built up this exposure time over 5--7 years.}
\label{fig:act_livetimes}
\end{figure*}

Sources were observed on a near-daily basis, but there are gaps of up to a few months due to inclement weather, telescope maintenance, hardware upgrades and seasonal visibility. As mm blazar jets exhibit strong variations on timescales of several months to years, this sampling cadence provides an excellent view of jet flaring activity. Distributions of source exposure times, after binning the data by 14 days, are shown in Figure~\ref{fig:act_livetimes}. This 14-day timescale is used later on in this work as a constant bin size to construct temporal neutrino expectations from measurement and from interpolation. 

\begin{figure*}[tb]
\centering
\includegraphics[width=\linewidth]{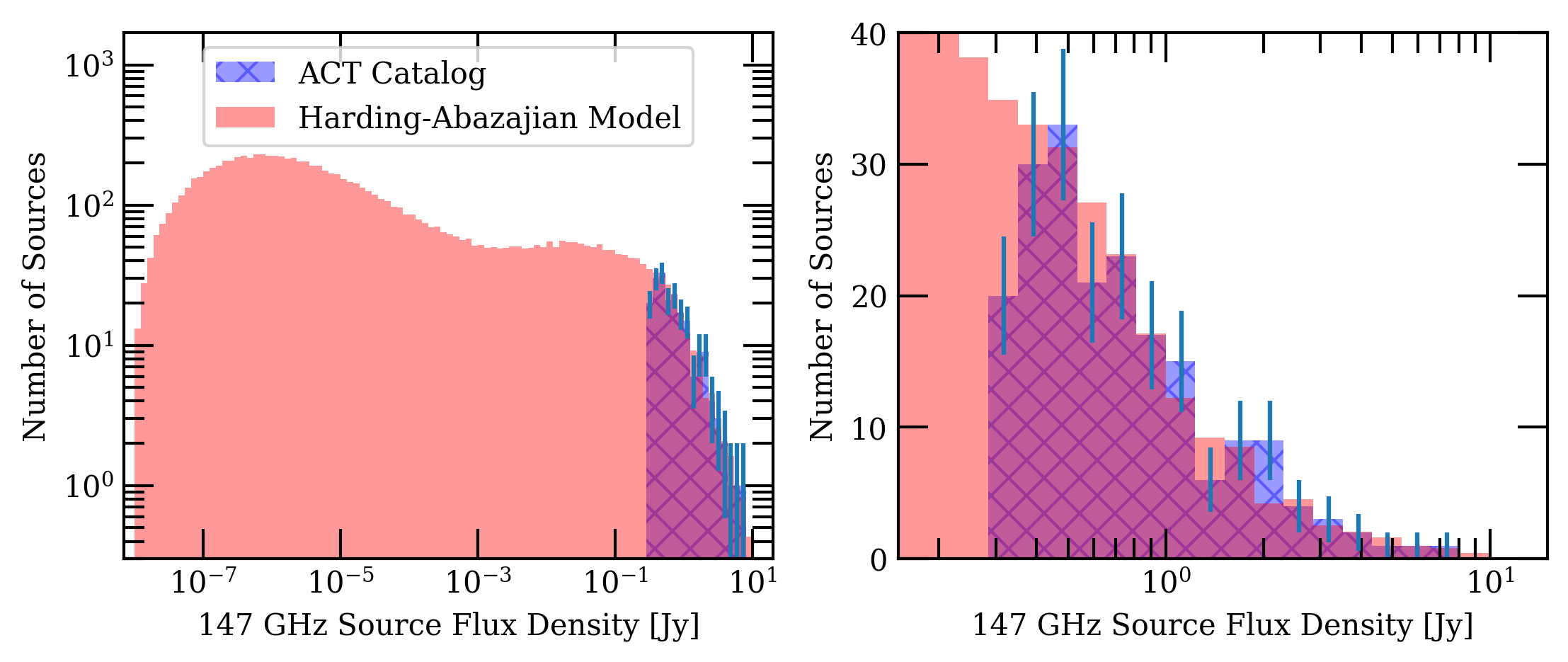}
\caption{Distributions indicating completeness of the ACT catalog relative to the tuned Harding--Abazajian evolutionary model. The left panel shows the entire distribution of the model, while the right panel shows the flux range in the ACT catalog. The distribution of average ACT blazar 147\,GHz flux densities is plotted on top of the source distribution predicted by the Harding--Abazajian model, after scaling for a mm-to-neutrino flux proportionality, as well as adjusting for the incomplete sky coverage of ACT (0.4) and for source number density as a function of flux; see text for details. The ACT catalog is flux-limited by the threshold of 500\,mJy at 95\,GHz that was used to select sources. While there is some slight bias in the catalog selection towards flaring, temporarily bright sources, the majority of blazar activity occurs on few-month-to-year timescales, so a representative description of average source brightness can still be provided with ACT's exposure time. The fraction of predicted flux from sources in the ACT catalog relative to the entire mm population is derived from this comparison. Accounting for sources outside of ACT's surveyed field, the final flux fraction or completeness associated with the ACT catalog is 29\%.}
\label{fig:act_completeness}
\end{figure*}

We estimate the flux completeness of the ACT AGN catalog relative to the expected distribution of blazar sources using the Harding--Abazajian luminosity-dependent blazar evolutionary model \citep{ Harding_2011, Harding_2012}. The fluxes of simulated source populations are generated as a function of redshift with FIRESONG  \citep{fire}. Because our analysis assumes a constant of proportionality between average flux density and neutrino intensity, the predicted distribution of arbitrary-unit source fluxes in the Harding--Abazajian model can be linearly scaled to match the measured ACT flux densities. The ACT AGN catalog was extracted from a footprint covering 40\% of the full sky. Thus, to allow for comparison with the ACT flux distribution, the predicted, full-sky population distribution is scaled by a factor of 0.40. This reduces the expected source number as a function of flux to reflect the limited solid angle viewed by ACT. Additionally, it is found that the resulting source number density as a function of flux is lower than that observed by ACT. The original Harding--Abazajian flux distribution is then adjusted to match the observed ACT blazar population. This final flux distribution is shown in solid red in Fig.~\ref{fig:act_completeness}. The distribution of 195 sources used from the ACT catalog are also shown in blue for comparison. The flux completeness of the ACT catalog specific to its surveyed solid angle is the integrated flux of the AGN in the catalog (the blue distribution of Fig.~\ref{fig:act_completeness}) relative to the integrated flux expected from the entire population (the red distribution). Relative to the full-sky, we find a final flux completeness of 29$\%$. Sources of the ACT catalog represent this fraction of the total flux expected from a comparable, complete, all-sky population.

\subsection{TXS 0506+056 as a Millimeter Source}\label{ssec:txs}

\begin{figure*}[tb]
\centering
\includegraphics[width=\linewidth]{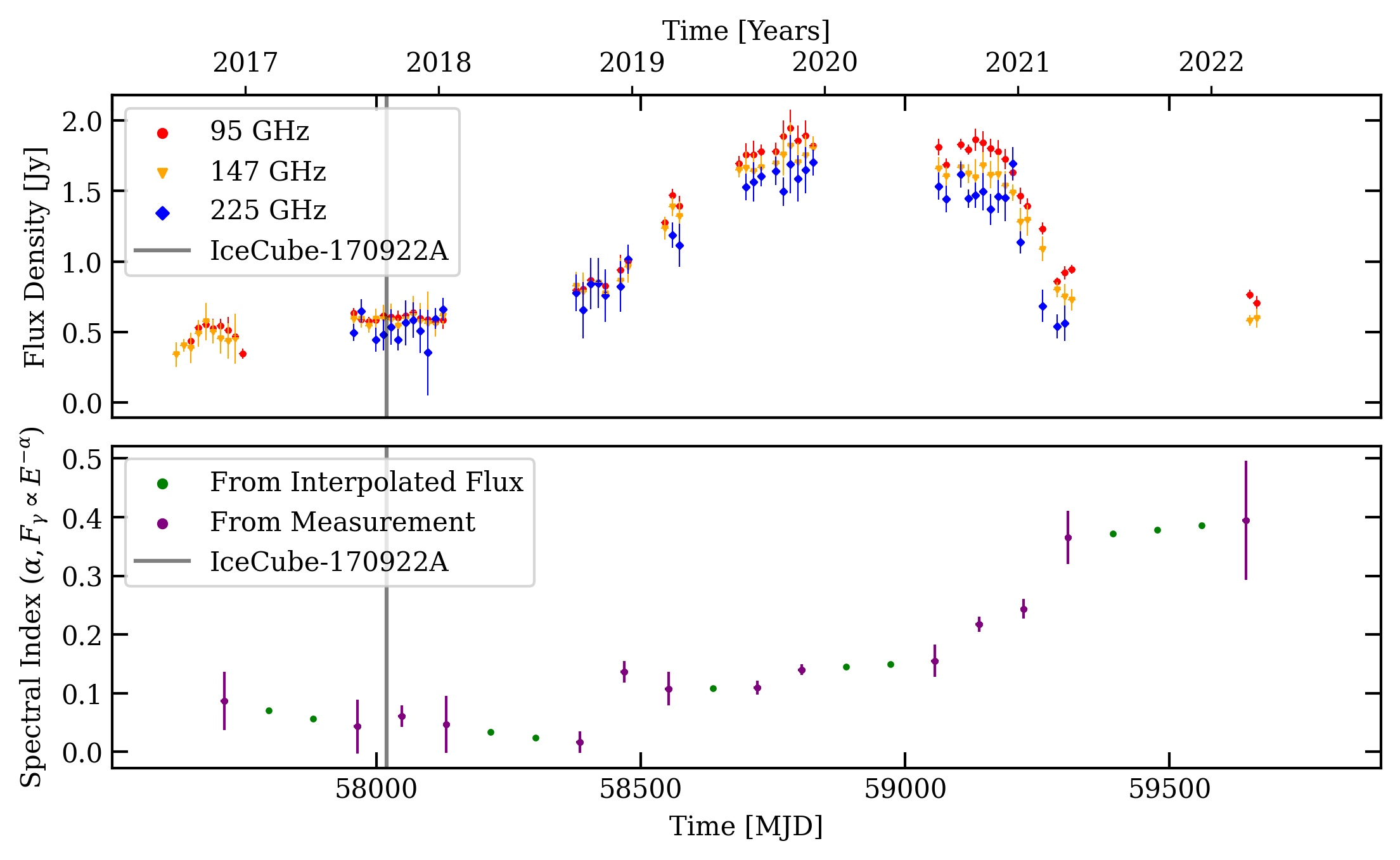}
\caption{Millimeter light curves and spectral index curve for TXS 0506+056. On top, the red, orange and blue data correspond to 95, 147 and 225\,GHz measurements averaged within 14-day bins. A substantial mm flare is observed between 2019 and 2021. On the bottom, a spectral index curve is determined from the interpolated 95 and 147\,GHz light curves with 84-day bins. A power-law relation is assumed between 95 and 147\,GHz flux densities. TXS~0506+056 shows a hardened state in the years surrounding the time of the IceCube $\sim$300 TeV track event. The power-law spectral index is at an absolute value near zero during this period, and slowly tapers off towards a value of 0.5. A spectral index of 0.5 to 0.7 is more typical of synchrotron spectra. The timescale of this variation is similar to that of the contemporaneous gamma-ray flare observed from the source \citep{Fermi_TXS_ATel}. The cadence of observations is representative of other sources within the catalog.}
\label{fig:txs_lightcurve}
\end{figure*}

TXS 0506+056 is a relatively bright mm-wavelength source with an average 147 GHz flux density of $\sim$1 Jy, in the top $27\%$ of the considered set of blazars. ACT observations of this source began in 2016, and include time periods prior to and during the 2017 IceCube alert event, IceCube-170922A \citep{Fermi_TXS_ATel}. A mm flare is observed between 2019 and 2022. The light curves in the three mm bands are shown in Fig.~\ref{fig:txs_lightcurve}. While TXS 0506+056 does not peak in flux density in 2017, a spectral hardening was noted, as seen in the `spectral index curve' of Fig.~\ref{fig:txs_lightcurve}. Here, `hardening' describes a relative increase in the intensity of higher-energy radiation. The spectral index is calculated assuming a power law between the 95 and 147\,GHz flux densities.

We consider that the substantial spectral hardening observed around the time of IceCube-170922A could represent an increase in higher-energy particle injection, and may be associated with neutrino production. This is shown in Fig.~\ref{fig:txs_lightcurve}. We also note the average mm spectral index of the source is exceptionally hard, lying in the top $\sim$5$\%$ of the spectral indices of
all sources within the catalog. 

\subsection{Potential Neutrino-Millimeter Temporal Correlations}\label{ssec:neutrino_mm_corr}

We consider three temporal models for a time-dependent correlation between mm and neutrino emission. In the first model, we assume a direct correlation between the neutrino intensity and the 147\,GHz flux density. Such a model suggests that neutrino production is always possible in mm-bright sources, but is modulated by mm-flaring activity. In this case, emission from the jet or particle loading site dominates our mm observations. The 147\,GHz band has the best livetime coverage and lowest uncertainty of the three bands included within the ACT catalog. These measurements are further binned into 14-day bins and averaged to minimize uncertainty. This timescale reflects a minimum expected period for major variations in source intensity. We also note that the 147\,GHz band is less likely to trace self-absorbed regions than the 95\,GHz band, which also has relatively low flux uncertainties. 

The second model is motivated by the expectation that non-variable mm emission from regions around the AGN may contribute to the observed mm flux density without tracing neutrino production. Specifically, cooled populations of electrons further down the jet may radiate and produce a steady-state SED for the source. Extended emission from the galaxy would also contribute to this quiescent baseline. These components would not trace particle loading at the base of the jet. It is therefore possible that the conditions required for neutrino production are not available in these regions. To trace only flaring activity from the jet base or other variable regions, our second model consists of correlating with the 147\,GHz mm light curve from which its minimum value has been subtracted, thereby assuming that there is no quiescent neutrino emission. The minimum used for the subtraction is obtained from the binned light curve described above.

Finally, for the third model, we note that while TXS 0506+056 did not show an apparent correlation between mm emission and the high-energy event IceCube-170922A, the source was in a significantly hardened state. We consider that this spectral hardening of the synchrotron spectrum may trace high-energy particle injection and neutrino production. For all sources, a similar spectral index curve is determined from interpolated 95 and 147\,GHz light curves. For each source, the top 32$\%$ of time in which the source is in its hardest state is chosen as a filter for the original 147\,GHz light curve. This threshold is phenomenologically motivated: it roughly selects the gamma-ray flaring period coincident with IceCube-170922A for TXS 0506+056. Neutrino emission is assumed here to correlate with the observed flux density only during these periods. We refer to this temporal model as the `index-filtered' model throughout this work.

\section{An All-Sky Selection of IceCube Events}\label{sec:icecube_data}

The IceCube South Pole Neutrino Observatory is a cubic kilometer of glacial ice instrumented with 5,160 photosensitive detectors. After a multi-year, staged deployment, eighty-six 2.5 kilometer-deep holes had been formed with hot water drilling by the end of 2010. A cable carrying power and communications hosts 60 detectors of approximate $\sim$17-meter equal spacing per hole between 1.5 and 2.5 kilometers in depth. Each sensor contains one downward-facing, hemispherical photomultiplier 10 inches in diameter. Observed photons or radiative emission associated with particle interactions in the ice are converted to a pulse of charge. The onboard electronics digitize triggered pulses and send them back to a collection house above the ice. Here, sufficient detections at multiple sensor locations satisfy a global event trigger. Data from the potential observed particle interaction are saved locally and communicated north via satellite \citep{icecube_technical}. 

Astrophysical neutrinos rarely interact on the path to Earth due to their low cross section. Within the higher-density earth and glacial ice, neutrinos interact through neutral- and charged-current channels. While a neutrino of lower energy exits in the neutral current interaction, the associated hadronic cascade leaves a point-like energy and light deposition within the detector. Such hadronic cascades are also produced in charged-current interactions, though the exiting charged lepton leaves an additional radiative signature within the detector. Muons produced in charged current interactions travel kilometer-scale distances without decaying, depositing radiation from Cherenkov, Bremsstrahlung, pair production and photo-nuclear interactions. The track-like signature and directionality of these interactions makes them ideal for point-source astronomy. Angular resolution improves with the event energy and amount of deposited light; such events have a ${\sim}0.3$-deg. angular uncertainty for energies in excess of 100\,TeV \citep{point_source_selection}. 

IceCube observes a substantial background from cosmic-ray air showers in the south. High-energy astrophysical cosmic rays interact in the upper atmosphere, creating showers of hadronic, leptonic, and radiative secondary particles. Secondary muons carry a large fraction of the primary cosmic-ray energy, and penetrate to the depth of IceCube as a highly-collimated bundle. This track-like signature is similar to that of the individual muons produced in charged-current neutrino interactions. Atmospheric electron and muon neutrinos are also produced as secondary particles. While atmospheric neutrinos form an additional background from all directions, the dominating muon background is reduced when observing tracks originating in the northern hemisphere (since the Earth provides a shield in this direction). Thus, IceCube generally has superior sensitivity to astrophysical point sources in the northern hemisphere relative to the southern hemisphere. 

We consider a selection of track-like events from the full sky. Data from April 2008 through May 2022 are used in the selection. Twelve years of these data were taken with the full detector array, and the initial two years were taken with the partially completed detector. In the south, a set of selections are used to remove likely atmospheric cosmic rays \citep{point_source_selection}. While this helps to improve sensitivity, a substantial background of atmospheric muons and neutrinos remains. 

\section{Methodology and Results}\label{sec:method_results}
\subsection{ A Search for Neutrino Emission from Individual ACT Blazars}\label{ssec:individual}

To test the three temporal models described in Sec.~\ref{ssec:neutrino_mm_corr}, direct correlation with the 147\,GHz light curve, the baseline-subtracted light curve, and periods filtered by spectral hardness, we first utilize spatial, energy and temporal information to search for neutrino emission from individual sources. Given an event of reconstructed declination and right ascension, $\delta_{i}$ and $\textrm{R.A.}_{i}$, respectively, or direction, $\vec{x} _{i}(\delta_{i}, \textrm{R.A.}_{i})$, angular uncertainty, $\sigma_{i}$, and a point source of known position, $\vec{x} _{s}(\delta, \textrm{R.A.})$, we express the spatial signal probability distribution function (PDF) as a 2D Gaussian on a sphere: 
\begin{linenomath*}
\begin{equation}
    S_{\mathrm{spat}}(\delta_{i}, \textrm{R.A.}_{i} |  \sigma _{i}) = \dfrac{1}{2 \pi \sigma _{i}^{2}}   \textrm{exp} \Bigg( - \dfrac{1}{2}  \bigg( \dfrac{  \vec{x} _{i} - \vec{x} _{s}  }{ \sigma _{i} }   \bigg)^{2} \Bigg). 
\end{equation}
\end{linenomath*}
Notably, the angular uncertainty of reconstructed IceCube events is at least an order of magnitude larger than localization errors associated with a source's cross match or the respective instrument uncertainties. 

Similarly, we define an energy signal PDF, $S_{\mathrm{ener}}(E_{i} | \delta, \gamma)$, as a function of reconstructed event energy, $E_{i}$, assuming a true power-law neutrino energy spectrum of index $\gamma$ from the source. Such a spectral shape is motivated by magnetic reconnection and shock-driven acceleration processes. This original power-law distribution may be parameterized as:
\begin{equation}
    \dfrac{dN_{\nu} }{ dE_{\nu} } = \Phi_{\nu_{l} + \overline{\nu}_{l} } \times \bigg( \dfrac{E_{\nu}  }{ \mathrm{100\,TeV} } \bigg)^{-\gamma}. 
\end{equation}
Here,  $E_{\nu}$ is the true neutrino energy. The normalization, $\Phi_{\nu_{l} + \overline{\nu}_{l}}$, encompasses the combined flux from both neutrinos and antineutrinos of a single flavor. We show results throughout this work as a function of this astrophysical, single-flavor power-law flux at 100 TeV. To determine the observed distribution, $S_{\mathrm{ener}}(E_{i} | \delta, \gamma)$, simulation is used to represent Earth propagation and reconstruction effects for the spectrum of a given source at declination, $\delta$. 

We also define a temporal signal PDF representing one of each of our three models, using ACT data binned into equal periods, averaged and normalized. This temporal signal PDF, $S_{\mathrm{temp}}(t_{i})$ is a function of event time, $t_{i}$. We note that in the linearly-interpolated and baseline-subtracted models, for time periods prior to the initial ACT observation and for periods after the last ACT observation, a constant value is assumed. Specifically, the average source flux density or average baseline-subtracted flux density is used for extrapolation. This assumption is equivalent to performing a joint time-independent analysis for these periods, where the observed duration provides an accurate representation of average mm intensity. 

Finally, for each of the temporal, spatial and energy PDFs, we define complementary background PDFs:  $B_{\mathrm{temp}}(t_{i})$, $B_{\mathrm{dec}}(\delta_{i})$, and $B_{\mathrm{ener}}(E_{i} | \delta_{i} )$. This background component represents events of atmospheric origin and any uncorrelated astrophysical events from alternate source populations. The temporal and spatial background PDFs, $B_{\mathrm{temp}}(t_{i})$ and $B_{\mathrm{dec}}(\delta_{i})$ assume a constant rate of atmospheric background specific to each declination band. This rate is also assumed constant as a function of right ascension, as represented by a factor of $1/(2\pi)$. The energy background PDF, $B_{\mathrm{ener}}(E_{i} | \delta_{i} )$, expresses the spectral shape of the background as a function of event declination. These three PDFs are estimated directly from data. 

With $N$ total events in the data selection, the likelihood of $n_{s}$ astrophysical neutrinos of spectral index, $\gamma$, originating from the source is, 
\onecolumngrid
\begin{multline}
    \mathcal{L} (n_{s}, \gamma) = \prod_{i=1}^{N}  \bigg[ \frac{n_{s}}{N} \cdot S_{\mathrm{temp}}(t_{i}) \cdot S_{\mathrm{spat}}(\delta_{i}, \mathrm{RA}_{i} | \sigma _{i}) \cdot S_{\mathrm{ener}}(E_{i} | \delta, \gamma) \textrm{ } + \\  \left(1 - \frac{n_{s}}{N}\right) \cdot \frac{1}{2\pi} \cdot B_{\mathrm{temp}}(t_{i}) \cdot B_{\mathrm{dec}}(\delta_{i}) \cdot B_{\mathrm{ener}}(E_{i} | \delta_{i} ) \bigg].
\end{multline}
\twocolumngrid

\noindent The likelihood test statistic, TS, can then be evaluated as a function of the likelihood-maximizing, best-fit signal parameters, $\hat{n}_{s}$ and $\hat{\gamma}$: 
\begin{equation}
\textrm{TS} = 2\left[\ln\mathcal{L}(n_{s} = \hat{n}_{s}, \gamma = \hat{\gamma}) - \ln\mathcal{L}( n_{s} = 0 )\right].
\end{equation}

Evaluation of this unbinned likelihood ratio follows from previous work \citep{Braun_2008}. As event rates are essentially constant as a function of right ascension, randomizing data in this dimension allows us to characterize search sensitivity subject to the dominating background component. We determine best-fit signal values for an ensemble of such event realizations. In practice, we optimize $n_{s}$ over a range of sampled $\gamma$ values. As a final step, the L-BFGS-B optimization routine is used for a multi-dimensional fit \citep{doi:10.1137/0916069, 10.1145/279232.279236, BFGS}. 

We inject signals of varied strength from simulations to determine search sensitivity and discovery potential flux. Given an astrophysical neutrino spectrum of specific spectral shape, the 90$\%$ sensitivity flux indicates the signal intensity at which the test-statistic distribution of an injected signal exceeds the median of the background test-statistic distribution 90$\%$ of the time. Similarly, the 5$\sigma$ discovery potential corresponds to the intensity at which the test-statistic distribution exceeds five standard deviations of the background test statistic distribution 50$\%$ of the time. Lastly, 90$\%$ confidence-limit upper limits may also be set by determining the flux intensity at which the test-statistic distribution exceeds the best-fit test statistic 90$\%$ of the time. In this work, we characterize our search sensitivities, discovery potentials and upper limits by the corresponding flux level at 100 TeV, $\Phi_{\nu_{l} + \overline{\nu}_{l}}$, assuming a specific spectral index, $\gamma$. 
In order to simplify the calculation of confidence intervals (as shown in Fig.~\ref{fig:txs_pos_spec}) we assume that the background test-statistic distribution is chi-squared distributed based on Wilks' theorem \citep{10.1214/aoms/1177732360}.

In our analysis of individual sources, we focus on only a subset of the most promising objects in order not to dilute a potential significant result with additional trials, as explained further in this section. The relative expected neutrino fluence is compared to the sensitivity flux for all 195 sources for each of the three temporal models. This fluence is proportional to the expected, model-dependent mm-fluence, equivalent to the time-integrated mm-flux density. An astrophysical neutrino spectrum of $\gamma = 2.0$ is assumed in all cases. This choice reflects the hard spectrum of candidate neutrino source, TXS~0506+056. 

\begin{figure*}[tb]
\centering
\includegraphics[width=\linewidth]{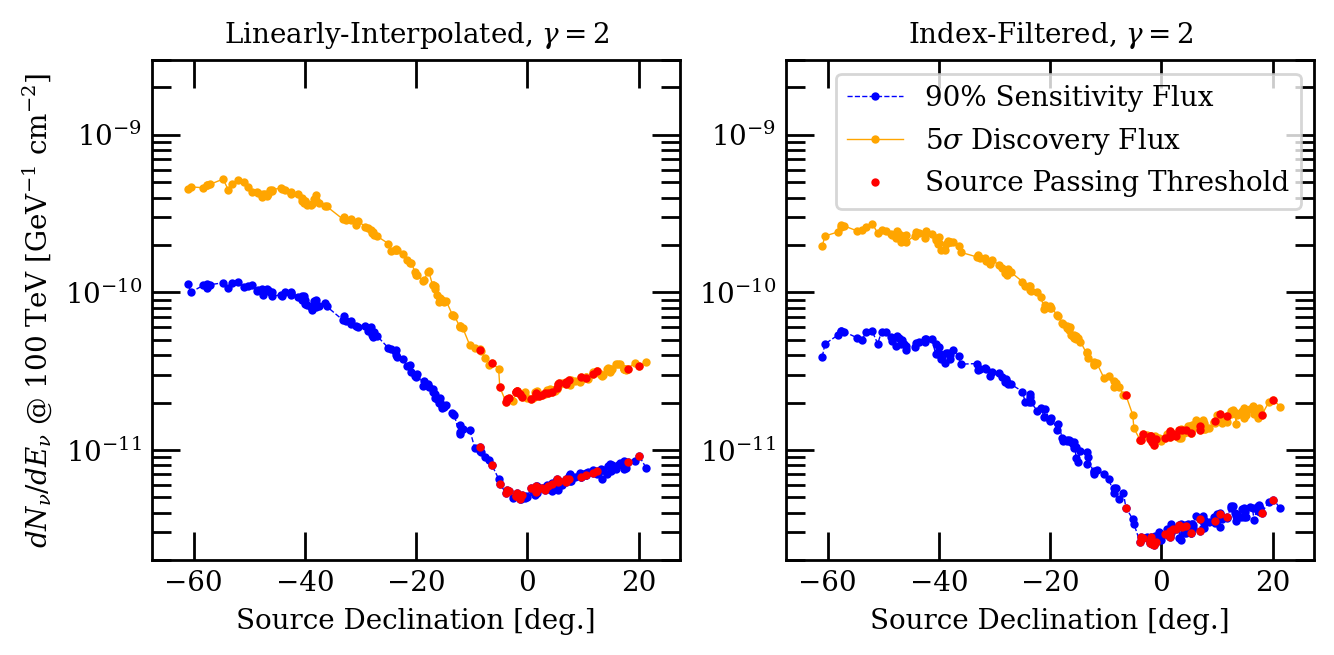}
\caption{ The sensitivity and discovery potential fluxes of IceCube searches for individual ACT blazars. Here, we show search sensitivity fluxes and 5$\sigma$ discovery potential fluxes for each individual source under two temporal model assumptions, a linear-interpolation of the light curve and a spectral-index-based filtering. The plotted flux level represents a single-flavor contribution from both neutrinos and antineutrinos. An injected power-law signal following an index of 2.0 is used. The stronger background rejection associated with index filtering corresponds to lower sensitivity and discovery potential fluxes. Sources indicated in red pass the criteria for analysis. At negative declinations, events originating from the southern sky have poorer sensitivity due to the large background of atmospheric cosmic-ray muons. Sensitivity also worsens with increasing declination in the north, where attenuation within the Earth diminishes the number of observed events, especially for neutrino energies beyond 100 TeV. }
\label{fig:sensitivity}
\end{figure*}

While an exact neutrino flux is not predicted for individual sources, we assume that the integrated mm fluence gives a description of relative neutrino intensity between different sources under a specific model. In this way, the ratio between the integrated mm model fluence and the required astrophysical neutrino flux for sensitivity can act as a statistic to rank the relevancy of each source. This statistic is expressed as:

\begin{multline}
    \textrm{Effective Weight} = \\ \dfrac{\textrm{Model Dependent Average Flux}}{\textrm{Source Sensitivity}}. 
\end{multline}

After ranking the sources under each model assumption, we select the set of sources with at least 10$\%$ the weight of the most highly-ranked source for analysis. The sensitivities of all individual and selected sources are shown for a correlation with the linearly-interpolated 147\,GHz light curve and with the index-filtered light curve in Fig.~\ref{fig:sensitivity}. This selection reduces the ultimate trials factor associated with the search. The final set of 45 sources selected for analysis based on any temporal model, their mm fluences and sky coordinates are listed in Table~\ref{tab:individual_sources} within the Appendix. A total of 33, 42 and 27 sources are selected for the linearly-interpolated, baseline-subtracted and index-filtered models, respectively. Results particular to each temporal model are described in the following section. 

\subsection{ Neutrino Emission from Individual ACT Blazars}\label{ssec:individual_result}

The results of the analysis for each temporal model are reported within the Appendix in Tables~\ref{tab:upper_limit_lin}, \ref{tab:upper_limit_baseline} and \ref{tab:upper_limit_index}. Sources with excesses corresponding to a minimum $2\sigma$ local significance are highlighted in Table~\ref{tab:indiv_summary}. The final trials-corrected $p$-value, taking into account all sources and all tested models, is provided for the most significant excess. The best-fit signal parameters for each source and temporal model are presented along with upper limits. There is a substantial correlation between the results of the linearly-interpolated and baseline-subtracted models. 

\begin{figure*}[tb]
\centering
\includegraphics[width=\linewidth]{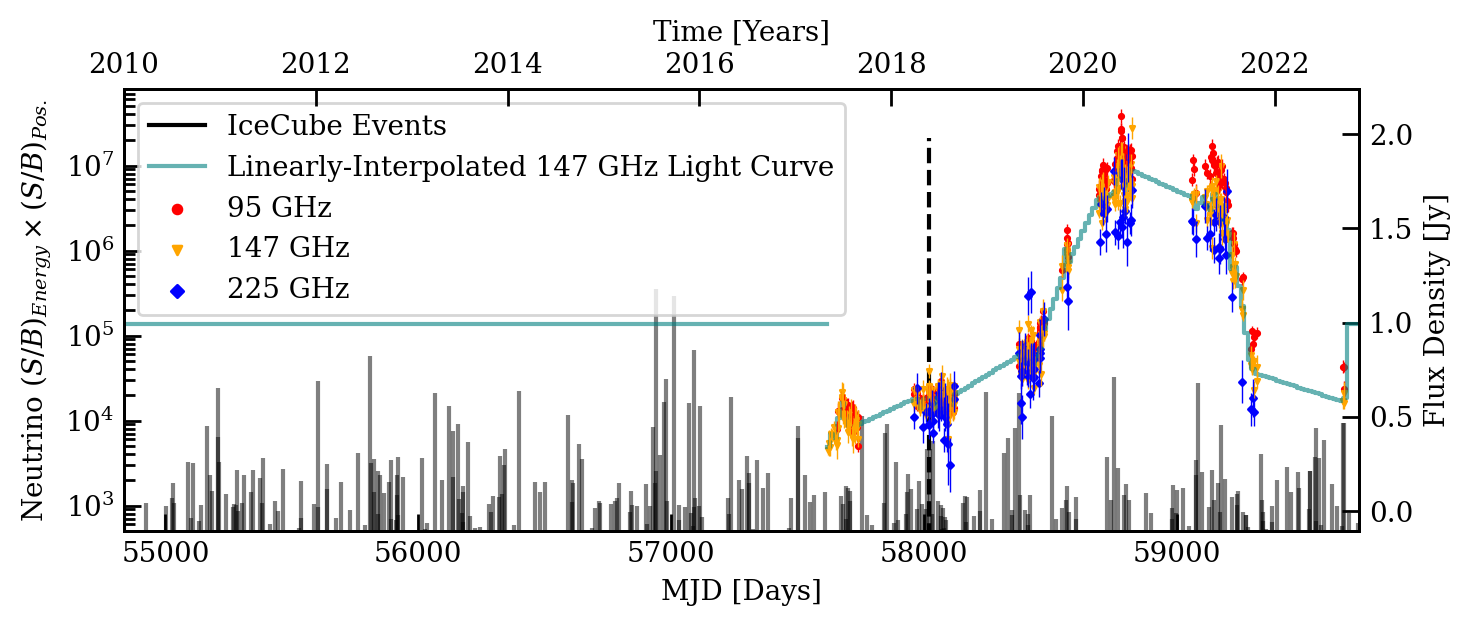}
\caption{Events contributing to the TXS 0506+056 excess. Here, we provide neutrino event energy and spatial weight in black as a function of time for events within 1.5 degrees of TXS 0506+056. Correlation with the linearly-interpolated 147\,GHz light curve (green line) is assumed. Prior to ACT observations, we extrapolate the mean 147\,GHz flux. The raw ACT data for 95, 147 and 225\,GHz are plotted for reference. The linearly-interpolated 147\,GHz light curve is also plotted with extrapolated coverage extended to periods before and after ACT observation. The alert event, IC-170922A, is indicated with a dashed line. An excess of events around 57000 MJD is also visible, corresponding to the 2014-2015 flare. There is not an obvious correlation between the mm flux density flare from 2019 to 2021 and any neutrino activity.  }
\label{fig:txs_events}
\end{figure*}

\begin{figure*}[tb]
\gridline{\fig{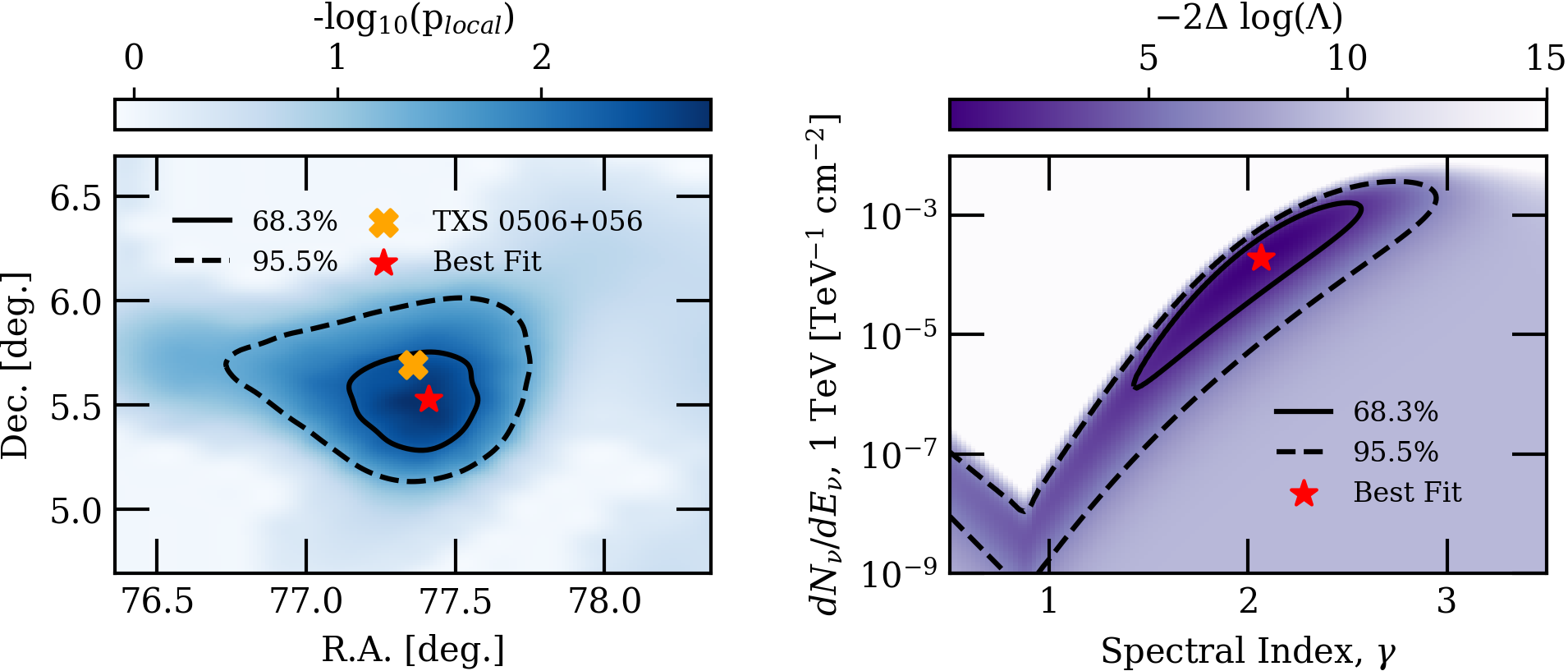}{
\textwidth}{  } }
\gridline{\fig{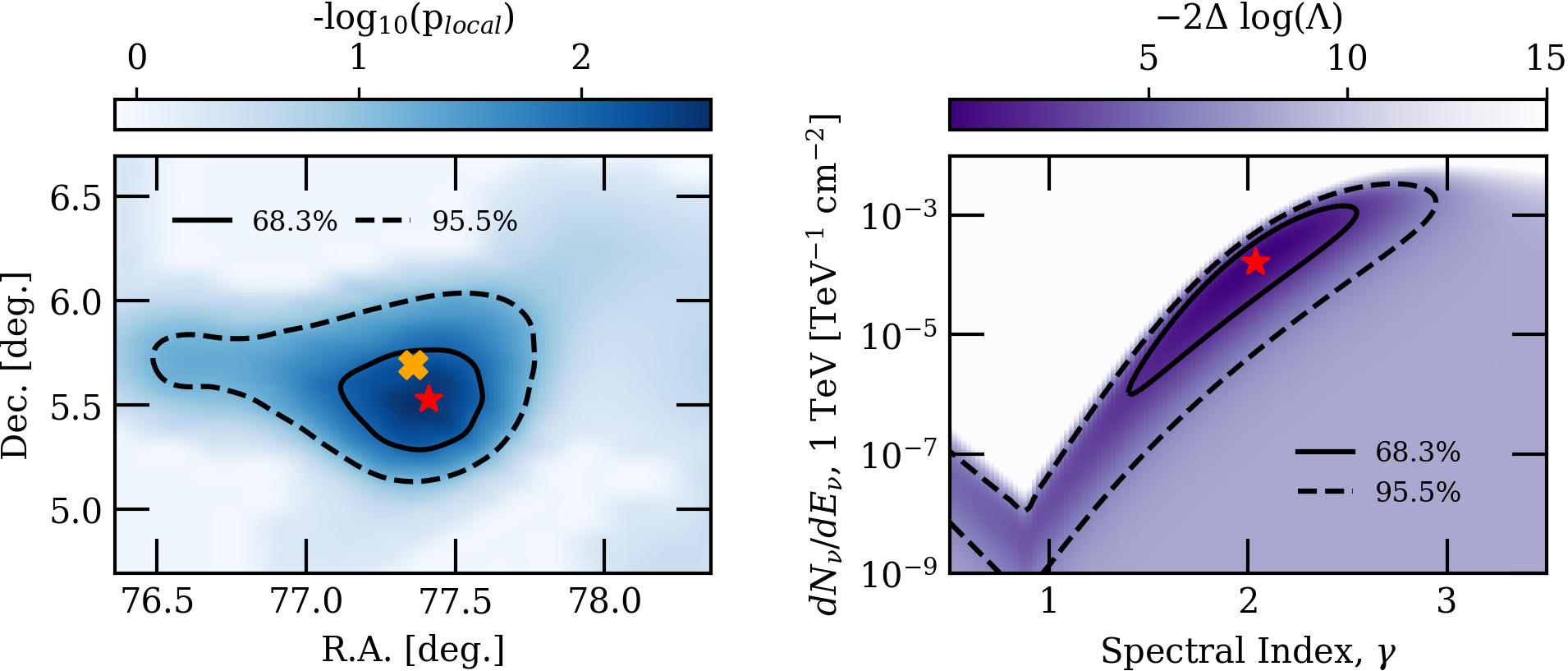} {\textwidth}{ } }
\caption{Positional and spectral scans of the IceCube TXS 0506+056 excess. Here, a positional and spectral scan of the TXS 0506+056 source location is shown in the top row assuming correlation with the linearly-interpolated 147\,GHz light curve and in the bottom row for the baseline-subtracted 147\,GHz light curve. On the left, the source likelihood is maximized as a function of assumed source right ascension and declination. The negative logarithm of the local $p$-value is plotted for each point. The known location of TXS 0506+056 is shown as an orange cross. On the right, the source likelihood is profiled as a function of signal intensity and power-law spectral index, $\gamma$. The logarithmic ratio between the best-fit likelihood and the likelihood of the selected signal parameters, $\Lambda$, is plotted. The best-fit locations for these spatial and energy signal parameters are pictured with a red star. 95.5$\%$ and 68.3$\%$ confidence regions are also shown, drawn from Wilks' theorem assuming two degrees of freedom.}
\label{fig:txs_pos_spec}
\end{figure*}

In all searches, TXS~0506+056 is the most significant source. We note that this result is likely driven by correlation with the known neutrino flare in 2014. In our analysis, this flare would have occurred during an earlier period prior to ACT observations, where an average flux density was assumed. This neutrino flare is actually known to have taken place during a radio quiet state, as established by observations from RATAN-600 \citep{Allakhverdyan_2023}. Additionally, the TXS~0506+056 alert event, IceCube-170922A, arrived during the mm-quiet state observed by ACT. We provide in Fig.~\ref{fig:txs_events} a comparison of spatial and energy neutrino event weights with the ACT mm light curves. It is worth noting that there is no clear increase in neutrino emission correlated with the observed mm flare between 2019 and 2021. We include likelihood scans of the spatial excess and spectral parameters of the source under the linearly-interpolated and baseline-subtracted temporal models in Fig.~\ref{fig:txs_pos_spec}.

Two other sources are observed in excess of a 2$\sigma$ local significance. The first, J1415+1320, is a BL Lacertae object (BL Lac) with potential evidence for gravitational lensing of the jet \citep{2017ApJ...845...90V, Vedantham_2017}. The second, 5BZB J0006-0623, is a low-synchrotron-peaked BL Lac \citep{2020NewA...7901393K}. We note that such excesses are neither significant nor unexpected given the number of trials performed in this analysis. After a correction for the number of trials, the global significance of the individual-source search is reduced to 0.8$\sigma$.

\begin{table*}[thb]
\begin{center}
\begin{tabular}{c c c c c c c c c }\hline \hline
Search & Source Name & Local Significance &  Post-Trials Significance & $\hat{n}_{s}$ & $\hat{\gamma}$ & $\phi_{90\%, \gamma = 2.0}$ & $\phi_{90\%, \gamma = 2.5}$  \\ \hline
\textbf{Linear-Interp.} & \textbf{TXS 0506+056} & \textbf{2.56$\sigma$}  & 0.8$\sigma$ & \textbf{9.94} & \textbf{2.06} & \textbf{20.08} & \textbf{50.13}  \\
Linear-Interp. & 5BZB J0006-0623 & 2.07$\sigma$ &   $-$ & 11.29 & 4.00 & 21.84 & 44.72 \\ \hline
\textbf{Shifted} & \textbf{TXS 0506+056} & \textbf{2.42$\sigma$} & $-$ &   \textbf{9.35} & \textbf{2.04} & \textbf{18.85} & \textbf{46.56} \\
Shifted & 5BZU J1415+1320 & 2.38$\sigma$ &  $-$ &  20.76 & 2.44 & 20.64 & 50.60  \\
\hline \hline 
\end{tabular} 
\caption{Summary of results from the individual source searches. Here, we provide best-fit results for sources of significance in excess of $2\sigma$ from the three catalog searches performed. Only under the linearly-interpolated and baseline-subtracted models are such signals found. Results of the most significant source of each temporal model are bolded. The best-fit results are provided for these sources with upper limits for power-law spectral indices of 2.0 and 2.5, as well as the local significance, and a final post-trials significance reflecting all sources tested under all temporal hypotheses. Upper limits are presented in the form, $dN_{\nu}/dE_{\nu}(\mathrm{100 \,TeV}) = \phi_{90\%} \, \times$ $10^{-12}$\,$\mathrm{GeV}^{-1}$ $\mathrm{cm}^{-2}$. The rate of a single-flavor flux, including both neutrinos and antineutrinos, is represented. }
\label{tab:indiv_summary}
\end{center}
\end{table*}

\subsection{ A Search for Neutrino Emission from the ACT Blazar Population}\label{ssec:population}

We also search for a combined signal from the entire population of ACT blazar AGN. In the event that individual sources are too faint to be observed, the cumulative emission may form a significant signal. A relative neutrino intensity is assumed between all sources, proportional to the integrated fluence expected for a specific temporal model. Additionally, a global neutrino spectrum is assumed for all sources, consisting of a power law following an index of $\gamma$. 

For a source population of size, $N_{src}$, we generalize an event's signal PDF to allow associations with all locations:
\onecolumngrid
\begin{equation}
\textrm{S}_{\mathrm{pop}}(\delta_{i}, \textrm{R.A.}_{i}, E_{i}, t_{i}) = \\ \dfrac{\sum\limits_{j=1}^{N_{\mathrm{src}}} w_{j} \cdot R(\delta_{j}, \gamma) \cdot S_{\mathrm{temp}, j}(t_{i}) \cdot S_{\mathrm{spat},j}(\delta_{i}, \textrm{R.A.}_{i} , \sigma_{i} ) \cdot S_{\mathrm{ener}, j}(E_{i} |\delta_{j}, \gamma) }{ \sum\limits_{j=1}^{N_{\mathrm{src}}} w_{j} \cdot R(\delta_{j}, \gamma) }. 
\end{equation}
\twocolumngrid
Here, $S_{\mathrm{spat},j}$ and $S_{\mathrm{ener},j}$ are the spatial signal and energy PDFs of the $j$th source and $w_{j}$ is a physically motivated weight controlling the relative amount of signal expected from a given source; in this case, $w_j$ is the time-integrated fluence from a specific temporal model. As detector effective area also varies as a function of source declination and neutrino energy, an additional weight is required to express this varying source acceptance, $R(\delta_{j}, \gamma)$.

The log-likelihood may then take the modified form,
\onecolumngrid
\begin{equation}
\ln(\mathcal{L}_{\mathrm{pop}}( n_{s}, \gamma )) = \sum_{i=1}^{N} \ln\Bigg[\dfrac{n_{s}}{N} \cdot  S_{\mathrm{pop}}(\delta_{i}, R.A._{i}, E_{i}, t_{i}) + \\  \bigg(1 -  \dfrac{n_{s}}{N} \bigg) \cdot  B_{\mathrm{temp}}(t_{i}) \cdot \dfrac{1}{2\pi} \cdot B_{\mathrm{dec}}(\delta_{i}) \cdot B_{\mathrm{ener}}(E_{i} | \delta_{i} ) \Bigg]. 
\end{equation}
\twocolumngrid

\subsection{ Neutrino Emission from the ACT Blazar Population}\label{ssec:population_result}

\begin{figure*}[tb]
\centering
\includegraphics[width=\linewidth]{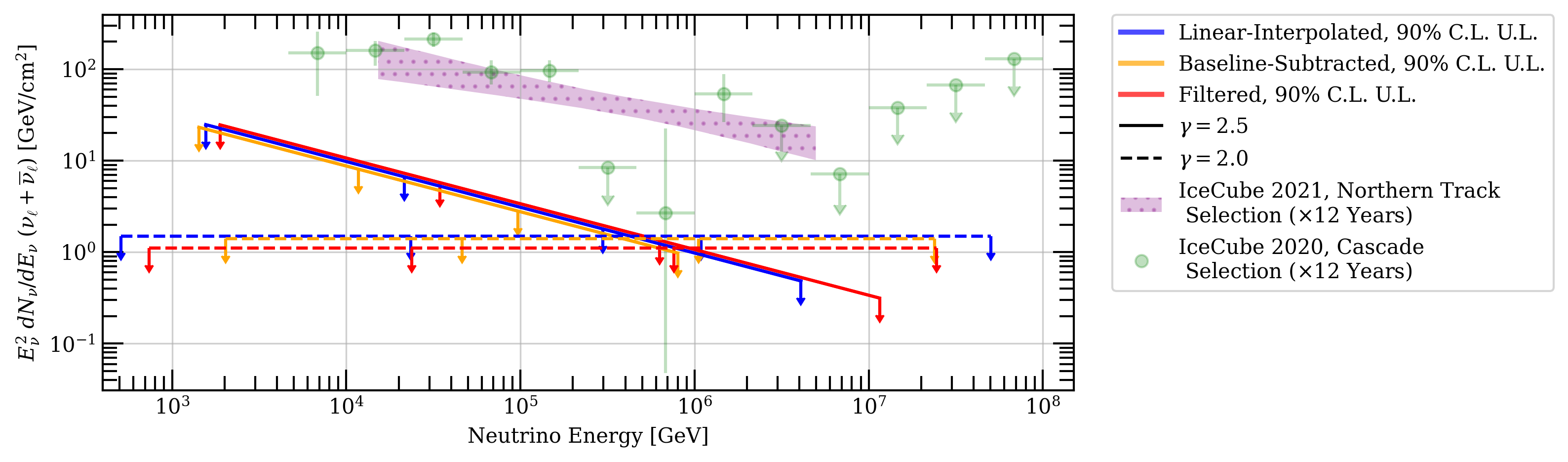}
\caption{Upper limits on the
contributions of the stacked population of ACT blazars to the
observed high-energy diffuse neutrino flux. Here, we show upper limits from the three stacked analyses performed, assuming neutrino spectra of indices 2.0 and 2.5. It should be noted that the upper limits for the linearly-interpolated and baseline-subtracted models for a spectral index of 2.0 are nearly overlapping. The flux intensity represents the cumulative flux from the stacked population, corrected to represent a complete set of sources. 90$\%$ sensitive energy ranges are shown. The purple butterfly band is taken from a recent measurement of the all-sky diffuse tau and electron flux and is shown for comparison \citep{PhysRevLett.125.121104}. Similarly, green data points are taken from an analysis of track-like events originating from the northern hemisphere \citep{2022ApJ...928...50A}. All pictured fluxes represent a single-flavor flux including both neutrinos and antineutrinos.}
\label{fig:stack_results}
\end{figure*}

We test each of three temporal models to search for emission from the set of 195 mm blazars. No significant signal is found under any model assumption, and we place upper limits. The best-fit results, local and trials-corrected significances are reported in Table~\ref{tab:stack_results}. After correcting for our three trials, the global significance of the test is 0.25$\sigma$. Fluxes and upper limits determined from this stacked analysis reflect only the combined contributions of sources within the ACT catalog. As the ACT catalog flux completeness has been determined, dividing such results by this factor will provide corresponding fluxes and limits from a complete, all-sky population of comparable blazars. These quantities are shown for comparison with previous measurements of the all-sky astrophysical diffuse neutrino flux in Fig.~\ref{fig:stack_results}. Including this correction for the geometric and redshift completeness of the ACT catalog, such a population of blazars is constrained to around 5$\%$ of the diffuse flux under the models assumed in this work. 

Energy ranges shown in Fig.~\ref{fig:stack_results} represent the 90$\%$ sensitive energy range of each presented temporal model and spectral index combination. A typical sensitive energy range for this analysis extends from 1\,TeV to 10\,PeV. The sensitivity depends on the energy spectra of the source population, source locations and time variability, among other characteristics, and we determine this sensitivity using simulations. To understand the dependence on energy, we adjust the lower or upper energy bound of injected source events upwards or downwards, respectively. When a 5$\%$ increase to the model sensitivity is determined for both the incremented lower and upper energy limits, this range is adopted as the sensitive energy range. Notably, these limits are fully derived from simulation, and the quoted energy ranges do not conflict with or relate to those resulting from recent diffuse analyses \citep{PhysRevLett.125.121104, 2022ApJ...928...50A}. 

\begin{table*}[tb]
\centering
\begin{tabular}{c c c c c c }\hline \hline
\textrm{\,\,\,\,\,\,} Search \textrm{\,\,\,\,\,\,} & \textrm{\,\,\,\,\,\,} TS \textrm{\,\,\,\,\,\,} & \textrm{\,\,\,\,\,\,} Pre-Trials p-value \textrm{\,\,\,\,\,\,} & \textrm{\,\,\,\,\,\,} Post-Trials p-value \textrm{\,\,\,\,\,\,} & \textrm{\,\,\,\,\,\,} $\phi_{90\%, \gamma = 2.0}$ \textrm{\,\,\,\,\,\,} & \textrm{\,\,\,\,\,\,} $\phi_{90\%, \gamma = 2.5}$ \textrm{\,\,\,\,\,\,}  \\ \hline
Linear-Interp. & 0.01 & 0.64 & $-$ & 4.41 & 9.10 \\ 
Shifted & 0.00 & 1.0 & $-$ & 4.13 & 8.13 \\ 
\textbf{Index-Filtered} & \textbf{2.16} & \textbf{0.14} & \textbf{0.25} & 3.26 & 9.94  \\ 
\hline \hline 
\end{tabular} 
\caption{Summary of results from the stacked population searches. In this table, we summarize the analysis results and significances for the three tested temporal models. Upper limits are presented in the form, $dN_{\nu}/dE_{\nu}(\mathrm{100 \,TeV}) = \phi_{90\%} \, \times$ $10^{-11}$\,$\mathrm{GeV}^{-1}$ $\mathrm{cm}^{-2}$. The most significant result comes from the index-filtered search. However, this mild excess is in-part related to the TXS 0506+056 alert event, which was included within the search. A significant result is not found from this stacked population of mm-bright blazars.}
\label{tab:stack_results}
\end{table*}

\section{Discussion and Conclusions}\label{sec:discussion}
\subsection{A Millimeter-Correlated Neutrino Flux}

Our analysis of individual mm-bright blazars and of the stacked population has not revealed a significant signal from the source class. While three sources, TXS 0506+056, 5BZB J0006-0623, and 5BZU J1415+1320, exceed a local significance of 2$\sigma$,  after a trials correction the result of this individual-source search is only significant at a level of $0.8\sigma$. Considering the entire population, neutrino production among blazars following a spectral index of 2.5 is not correlated with mm-synchrotron activity at a level of ${\sim}5\%$ of the all-sky astrophysical neutrino flux. We note that previous analyses of public data conflict with this result. Specifically, a correlation with radio-bright blazar cores has been suggested \citep{plavin2020, radio_2021, owens_metsahovi, Plavin_2023}. These analyses have generally focused on correlations with available, longer-wavelength radio emission, and have proposed that a superior tracer of time-dependent neutrino production would come from the unabsorbed mm wavelengths \citep{plavin2020}. The ACT mm-blazar catalog allows for this improved analysis of less-absorbed mm activity. The increased IceCube data livetime and the improved analysis methods of this work utilize event observables (reconstructed energy, direction and time) to provide a maximally sensitive search. As these millimeter wavelengths are generally accepted as a preferred tracer of radio--mm synchrotron activity from the blazar core, it is more likely that radio--mm blazar emission is not an important tracer of neutrino production.

A neutrino cascade interaction of ${\sim}225$\,TeV in energy was observed by the Baikal Gigaton Volume Detector (Baikal-GVD) from the vicinity of TXS~0506+056 in April 2021. Due to large spatial uncertainties, the event was not significantly correlated. However, the arrival of the event during a radio flare from TXS~0506+056 was noted as potential evidence in favor of radio-correlated neutrino production \citep{Allakhverdyan_2023}. We do not see any indication of new neutrino emission from the source during its observed mm flare from 2019 to 2021. Neutrino emission from TXS 0506+056 does not appear to trace its synchrotron intensity. 

While this work has constrained a general contribution from synchrotron-bright blazars, such synchrotron flares may trace or even facilitate neutrino production for a subset of sources. This relationship depends highly on the injected electron and proton populations, Compton-dominance of the injection zone, and whether the region is sufficiently magnetically dominated such that the energy output in synchrotron is significantly detected over the steady-state SED. Modeling of these flares may require treatment of each individual blazar and its unique jet and loading conditions. Based on this work, we can only infer that the general synchrotron-bright flare is either proton-light or does not offer conditions for efficient neutrino production.  We have also placed upper limits on some of the brightest individual mm blazars, like 3C 273 (5BZQ J1229+0203). Future follow-ups will benefit greatly from the continued monitoring of blazar mm light curves with the Simons Observatory \citep{so-collaboration/2019,so_collaboration:2025} and CMB-S4 \citep{cmbs4-collaboration/2021}.
 
\subsection{Blazars as Neutrino Sources}
While we find no evidence for a neutrino correlation with blazar-synchrotron activity, blazars are still expected to contribute as an astrophysical neutrino source class. Blazars offer some of the most powerful jets in the universe, sources of accelerated electrons and protons, and fields of target photons \citep{hovatta2020relativisticjetsblazars}. 

The sources TXS~0506+056, PKS~1424+240 and GB6 J1542+6129 share similar radiative spectral features and have been associated with excesses of astrophysical neutrinos. Specifically, they have hard, bright synchrotron spectra with a high peak in the optical or UV range. While traditionally classified as BL Lac objects, recent work has suggested broad line regions, emitting regions within one parsec of the central black hole, may be present. The broad optical emission lines expected from these zones would be subdominant to the bright synchrotron spectra. These photon fields appear boosted in the frame of the jet, creating targets for potential p-gamma collisions. The combination of high acceleration powers for both electrons and protons and these boosted photon populations might make this specific class of ``masquerading BL Lacs'' efficient neutrino emitters \citep{Padovani_2019, Padovani_2022}. 

Alternate modeling has also proposed a supermassive black hole binary system as an explanation for the TXS~0506+056 jet activity \citep{de_Bruijn_2020, Tjus_2022}. A precessing jet can be linked to the binary motion, ultimately modulating the intensity of observed neutrinos. If this model correctly explained neutrino emission in 2014--2015 and in 2017, a new flare would have been expected between 2019 and 2021, when the mm brightness increases (see Fig.~\ref{fig:txs_lightcurve}). Such activity is not observed in the most recent IceCube data. 

More specialized modeling and selections may be required to detect neutrino emission from blazars and to probe specific production scenarios. The presence of a coronal target photon field, high-peaked synchrotron spectra, broad lines, or other sources of target photons may indicate an efficient production environment. Selections based on these observable properties of the jet may have sensitivity to emission from a subset of blazars and are still viable. Alternatively, an excess of gamma rays from the Bethe Heitler process or pion decay may indicate the presence of proton populations. In general, difficulty in forming these selections comes from either a lack of knowledge about relative proton and electron distributions or about target photon fields. Still, it is this model development and testing which will help to constrain these physical quantities like proton luminosity in future work.

\section*{Acknowledgments}
The IceCube collaboration acknowledges the significant contribution to this manuscript from Alina Kochocki. The authors gratefully acknowledge the support for IceCube from the following agencies and institutions: USA $-$ U.S. National Science Foundation-Office of Polar Programs, U.S. National Science Foundation-Physics Division, U.S. National Science Foundation-EPSCoR, U.S. National Science Foundation-Office of Advanced Cyberinfrastructure, Wisconsin Alumni Research Foundation, Center for High Throughput Computing (CHTC) at the University of Wisconsin-Madison, Open Science Grid (OSG), Partnership to Advance Throughput Computing (PATh), Advanced Cyberinfrastructure Coordination Ecosystem: Services $\&$\ Support (ACCESS), Frontera computing project at the Texas Advanced Computing Center, U.S. Department of Energy-National Energy Research Scientific Computing Center, Particle astrophysics research computing center at the University of Maryland, Institute for Cyber-Enabled Research at Michigan State University, Astroparticle physics computational facility at Marquette University, NVIDIA Corporation, and Google Cloud Platform; Belgium $-$ Funds for Scientific Research (FRS-FNRS and FWO), FWO Odysseus and Big Science programmes, and Belgian Federal Science Policy Office (Belspo); Germany $-$ Bundesministerium f{\"u}r Bildung und Forschung (BMBF), Deutsche Forschungsgemeinschaft (DFG), Helmholtz Alliance for Astroparticle Physics (HAP), Initiative and Networking Fund of the Helmholtz Association, Deutsches Elektronen Synchrotron (DESY), and High Performance Computing cluster of the RWTH Aachen; Sweden $-$ Swedish Research Council, Swedish Polar Research Secretariat, Swedish National Infrastructure for Computing (SNIC), and Knut and Alice Wallenberg Foundation; European Union $-$ EGI Advanced Computing for research; Australia $-$ Australian Research Council; Canada $-$ Natural Sciences and Engineering Research Council of Canada, Calcul Qu\'ebec, Compute Ontario, Canada Foundation for Innovation, WestGrid, and Digital Research Alliance of Canada; Denmark $-$ Villum Fonden, Carlsberg Foundation, and European Commission; New Zealand $-$ Marsden Fund; Japan $-$ Japan Society for Promotion of Science (JSPS) and Institute for Global Prominent Research (IGPR) of Chiba University; Korea $-$ National Research Foundation of Korea (NRF); Switzerland $-$ Swiss National Science Foundation (SNSF).

The ACT collaboration acknowledges the significant contribution to this manuscript from Adam Hincks, Xiaoyi Ma, Cristian Vargas and Carlos Herv\'ias-Caimapo. The authors thank Suzanne Staggs for feedback on the manuscript.  Support for ACT was provided by the U.S.~National Science Foundation through awards AST-0408698, AST-0965625, and AST-1440226 for the ACT project, as well as awards PHY-0355328, PHY-0855887 and PHY-1214379. Funding was also provided by Princeton University, the University of Pennsylvania, and a Canada Foundation for Innovation (CFI) award to UBC. ACT operated in the Parque Astron\'omico Atacama in northern Chile under the auspices of the Agencia Nacional de Investigaci\'on y Desarrollo (ANID). The development of multichroic detectors and lenses was supported by NASA grants NNX13AE56G and NNX14AB58G. Detector research at NIST was supported by the NIST Innovations in Measurement Science program. Computing for ACT was performed using the Princeton Research Computing resources at Princeton University, the National Energy Research Scientific Computing Center (NERSC), and the Niagara supercomputer at the SciNet HPC Consortium. SciNet is funded by the CFI under the auspices of Compute Canada, the Government of Ontario, the Ontario Research Fund-Research Excellence, and the University of Toronto. We thank the Republic of Chile for hosting ACT in the northern Atacama, and the local indigenous Licanantay communities whom we follow in observing and learning from the night sky. This work was supported by a grant from the Simons Foundation (CCA 918271, PBL). ADH acknowledges support from the Sutton Family Chair in Science, Christianity and Cultures, from the Faculty of Arts and Science, University of Toronto, and from the Natural Sciences and Engineering Research Council of Canada (NSERC) [RGPIN-2023-05014, DGECR-2023-00180]. CHC acknowledges ANID FONDECYT Postdoc Fellowship 3220255 and BASAL CATA FB210003. CS acknowledges support from the Agencia Nacional de Investigaci\'on y Desarrollo (ANID) through Basal project FB210003.



\appendix
\section{Tables of Sources and Results}\label{app:tables}

In this appendix we provide complete tables of the ACT sources used for individual searches (Table~\ref{tab:individual_sources}), as well as the list of sources selected for each of the linearly-interpolated, baseline-subtracted and index-filtered individual searches (Tables~\ref{tab:upper_limit_lin}, \ref{tab:upper_limit_baseline} and \ref{tab:upper_limit_index}, respectively).

\begin{table}[tb]
\centering
\begin{tabular}{c c c c c}\hline \hline
Source Name & \phantom{gg } Blazar Class \phantom{gg} &  \phantom{gg} Average 147\,GHz Flux Density [Jy]  \phantom{gg} &  \phantom{gg} R.A. [deg.]  \phantom{gg} &  \phantom{gg} Dec. [deg.] \phantom{} \\ \hline
5BZQ J1229+0203 & FSRQ & 7.67 & 187.28 & 2.05 \\
5BZQ J2232+1143 & FSRQ & 3.63 & 338.15 & 11.73 \\
5BZB J0854+2006 & BL Lac & 3.60 & 133.70 & 20.11 \\
5BZU J1058+0133 & BCU & 3.41 & 164.62 & 1.57 \\
5BZQ J1833-2103 & FSRQ & 3.14 & 278.42 & -21.06 \\
5BZQ J0423-0120 & FSRQ & 2.77 & 65.82 & -1.34 \\
5BZU J0433+0521 & BCU & 2.54 & 68.30 & 5.35 \\
5BZB J0006-0623 & BL Lac & 2.44 & 1.56 & -6.39 \\
5BZU J0725-0054 & BCU & 2.36 & 111.46 & -0.92 \\
5BZQ J1743-0350 & FSRQ & 2.18 & 266.00 & -3.83 \\
5BZB J1751+0939 & BL Lac & 2.01 & 267.89 & 9.65 \\
5BZQ J1549+0237 & FSRQ & 1.94 & 237.37 & 2.62 \\
5BZQ J0510+1800 & FSRQ & 1.93 & 77.51 & 18.01 \\
5BZQ J0739+0137 & FSRQ & 1.67 & 114.83 & 1.62 \\
5BZQ J2229-0832 & FSRQ & 1.64 & 337.42 & -8.55 \\
5BZQ J2148+0657 & FSRQ & 1.57 & 327.02 & 6.96 \\
5BZQ J0501-0159 & FSRQ & 1.57 & 75.30 & -1.99 \\
5BZQ J1504+1029 & FSRQ & 1.54 & 226.10 & 10.49 \\
5BZQ J0108+0135 & FSRQ & 1.46 & 17.16 & 1.58 \\
5BZQ J0224+0659 & FSRQ & 1.23 & 36.12 & 6.99 \\
5BZQ J2101+0341 & FSRQ & 1.16 & 315.41 & 3.69 \\
5BZQ J2136+0041 & FSRQ & 1.15 & 324.16 & 0.70 \\
5BZB J0831+0429 & BL Lac & 1.13 & 127.95 & 4.49 \\
5BZQ J0339-0146 & FSRQ & 1.10 & 54.88 & -1.78 \\
5BZQ J2123+0535 & FSRQ & 1.08 & 320.94 & 5.59 \\
5BZB J0238+1636 & BL Lac & 1.08 & 39.66 & 16.62 \\
5BZB J2134-0153 & BL Lac & 1.06 & 323.54 & -1.89 \\
5BZQ J0532+0732 & FSRQ & 1.05 & 83.16 & 7.55 \\
5BZQ J0750+1231 & FSRQ & 1.01 & 117.72 & 12.52 \\
5BZB J0509+0541 & BL Lac & 0.99 & 77.36 & 5.69 \\
5BZQ J2225-0457 & FSRQ & 0.91 & 336.45 & -4.95 \\
5BZQ J2218-0335 & FSRQ & 0.89 & 334.72 & -3.59 \\
5BZB J0825+0309 & BL Lac & 0.87 & 126.46 & 3.16 \\
5BZQ J2301-0158 & FSRQ & 0.85 & 345.28 & -1.97 \\
5BZB J0217+0837 & BL Lac & 0.83 & 34.32 & 8.62 \\
5BZQ J2323-0317 & FSRQ & 0.77 & 350.88 & -3.28 \\
5BZQ J1224+2122 & FSRQ & 0.74 & 186.23 & 21.38 \\
5BZQ J0217+0144 & FSRQ & 0.70 & 34.45 & 1.75 \\
5BZQ J1224+0330 & FSRQ & 0.66 & 186.22 & 3.51 \\
5BZQ J1222+0413 & FSRQ & 0.63 & 185.59 & 4.22 \\
5BZU J1415+1320 & BCU & 0.58 & 214.00 & 13.34 \\
5BZQ J2327+0940 & FSRQ & 0.56 & 351.89 & 9.67 \\
5BZQ J1028-0236 & FSRQ & 0.52 & 157.14 & -2.62 \\
5BZB J0811+0146 & BL Lac & 0.48 & 122.86 & 1.78 \\
5BZQ J0839+0104 & FSRQ & 0.46 & 129.96 & 1.07 \\
\hline \hline 
\end{tabular} 
\caption{ACT sources unblinded in individual IceCube searches. Here, we provide information on the combined set of blazars analyzed as individual sources. The BZCAT name, classification, average 147\,GHz flux density, and equatorial coordinates are provided. Here, `BCU' (blazar class unknown) is taken to represent those sources classified with `blazar uncertain type'. Source coordinates are adopted from either the VizieR database, the NVSS databases, or the AT20G Australia Telescope 20\,GHz survey catalog through a cross-match process with ACT localizations. We note that the source, TXS 0506+056, is listed as 5BZB J0509+0541 in this table and Tables \ref{tab:upper_limit_lin}, \ref{tab:upper_limit_baseline} and \ref{tab:upper_limit_index}. Each source is tested under at least one temporal assumption -- linear-interpolation, baseline-subtraction or index-filtering. }
\label{tab:individual_sources}
\end{table}

\begin{table}
\centering
\begin{tabular}{c c c c c c c}\hline \hline
Source Name & Model Rank (Linearly-Interpolated) & $-$log$_{10}(p_{\mathrm{local}}$) & $\hat{n}_{s}$ & $\hat{\gamma}$ & $\phi_{90\%, \gamma = 2.0}$ & $\phi_{90\%, \gamma = 2.5}$    \\ \hline
5BZQ J1229+0203 & 1.00 & 0.65 & 4.01 & 2.2 & 8.59 & 22.93 \\
5BZU J1058+0133 & 0.47 & 0.00 & 0.00 & 4.00 & 5.05 & 11.99 \\
5BZQ J0423-0120 & 0.43 & 0.21 & 2.81 & 4.00 & 4.8 & 11.61 \\
5BZQ J2232+1143 & 0.38 & 0.53 & 1.04 & 1.48 & 9.61 & 24.66 \\
5BZU J0725-0054 & 0.34 & 0.64 & 15.79 & 4.00 & 7.91 & 20.55 \\
5BZQ J1743-0350 & 0.31 & 0.32 & 0.65 & 4.00 & 5.83 & 13.50 \\
5BZB J0854+2006 & 0.30 & 0.35 & 3.36 & 2.63 & 9.90 & 20.30 \\
5BZU J0433+0521 & 0.29 & 0.58 & 16.80 & 3.87 & 9.67 & 24.50 \\
5BZQ J1549+0237 & 0.25 & 0.25 & 1.50 & 3.87 & 5.71 & 14.25 \\
5BZQ J0501-0159 & 0.23 & 0.49 & 12.96 & 4.00 & 6.75 & 17.55 \\
\textbf{5BZB J0006-0623} & \textbf{0.23} & \textbf{1.74} & \textbf{11.29} & \textbf{4.00} & \textbf{21.84} & \textbf{44.72} \\
5BZB J1751+0939 & 0.23 & 0.33 & 3.08 & 2.68 & 7.61 & 17.07 \\
5BZQ J0739+0137 & 0.21 & 0.51 & 2.73 & 2.09 & 8.02 & 18.83 \\
5BZQ J0108+0135 & 0.20 & 0.20 & 0.75 & 3.06 & 5.23 & 12.45 \\
5BZQ J2148+0657 & 0.19 & 0.00 & 0.00 & 4.00 & 5.53 & 16.42 \\
5BZQ J0510+1800 & 0.17 & 0.00 & 0.00 & 3.25 & 6.93 & 18.33 \\
5BZQ J1504+1029 & 0.17 & 0.61 & 8.88 & 2.52 & 10.16 & 24.77 \\
5BZB J2134-0153 & 0.16 & 0.44 & 10.78 & 2.95 & 6.39 & 16.25 \\
5BZQ J0339-0146 & 0.15 & 0.52 & 10.48 & 4.00 & 7.62 & 18.57 \\
5BZQ J2136+0041 & 0.15 & 0.80 & 18.36 & 3.75 & 9.52 & 22.88 \\
5BZQ J2101+0341 & 0.15 & 0.00 & 0.00 & 4.00 & 5.37 & 13.14 \\
5BZQJ0224+0659 & 0.14 & 0.00 & 0.00 & 3.00 & 7.16 & 15.93 \\
5BZB J0831+0429 & 0.14 & 0.58 & 13.25 & 2.80 & 8.74 & 21.60 \\
5BZQ J2123+0535 & 0.13 & 0.00 & 0.00 & 4.00 & 6.03 & 13.24 \\
5BZQ J2301-0158 & 0.13 & 0.00 & 0.00 & 3.00 & 4.97 & 12.42 \\
5BZQ J0532+0732 & 0.12 & 0.00 & 0.00 & 4.00 & 6.83 & 17.87 \\
5BZQ J2229-0832 & 0.12 & 0.00 & 0.00 & 3.00 & 11.08 & 17.90 \\
5BZQ J2218-0335 & 0.12 & 0.37 & 2.27 & 4.00 & 6.54 & 13.92 \\
\textbf{5BZB J0509+0541} & \textbf{0.12} & \textbf{2.42} & \textbf{9.94} & \textbf{2.06} & \textbf{20.08} & \textbf{50.13} \\
5BZB J0825+0309 & 0.12 & 0.79 & 20.31 & 3.25 & 9.72 & 24.28 \\
5BZQ J2225-0457 & 0.11 & 0.00 & 0.0 & 2.25 & 5.47 & 14.27 \\
5BZQ J2323-0317 & 0.11 & 0.30 & 3.31 & 3.31 & 5.44 & 13.67 \\
5BZQ J0750+1231 & 0.10 & 0.86 & 21.47 & 3.22 & 13.19 & 34.06 \\
\hline \hline 
\end{tabular} 
\caption{ACT sources selected for linearly-interpolated individual searches. In this table we report best-fit results and 90$\%$ upper limits for sources tested with the linearly-interpolated temporal model. This model reflects the assumption that neutrino flux increases proportional to its mm flux density. The two sources in excess of $2\sigma$ are bolded within the table. Sources are listed in order of their model rank. The effective source weight statistic is determined from the model-dependent average flux and search sensitivity. This model rank is reported as a relative fraction of the maximum source statistic for this set. The negative logarithm of the local p-value of the search and best-fit signal parameters, $\hat{n}_{s}$ and $\hat{\gamma}$ are reported. Lastly, upper limits are provided for an assumed neutrino spectrum of index 2.0 and 2.5. Here, upper limits are presented in the form, $dN_{\nu}/dE_{\nu}(\mathrm{100 \,TeV}) = \phi_{90\%} \, \times$ $10^{-12}$\,$\mathrm{GeV}^{-1}$ $\mathrm{cm}^{-2}$, and represent the rate of neutrinos and antineutrinos of a single flavor.
 }
\label{tab:upper_limit_lin}
\end{table}

\begin{table}[tb]
\centering
\begin{tabular}{c c c c c c c}\hline \hline
Source Name & Model Rank (Baseline-Subtracted) & $-$log$_{10}(p_{\mathrm{local}}$) & $\hat{n}_{s}$ & $\hat{\gamma}$ & $\phi_{90\%, \gamma = 2.0}$ & $\phi_{90\%, \gamma = 2.5}$   \\ \hline
5BZQ J1229+0203 & 1.00 & 0.71 & 3.62 & 2.14 & 8.78 & 21.27 \\
5BZQ J0423-0120 & 0.83 & 0.22 & 3.26 & 4.00 & 4.76 & 11.30 \\
5BZQ J2232+1143 & 0.69 & 0.48 & 0.96 & 1.49 & 8.72 & 22.91 \\
5BZU J1058+0133 & 0.56 & 0.24 & 1.99 & 4.00 & 4.97 & 12.11 \\
5BZU J0433+0521 & 0.43 & 0.61 & 16.77 & 3.89 & 9.69 & 24.48 \\
5BZB J0854+2006 & 0.42 & 0.40 & 5.46 & 2.77 & 10.59 & 21.93 \\
5BZQ J1743-0350 & 0.42 & 0.00 & 0.00 & 4.00 & 5.22 & 12.79 \\
5BZQ J0739+0137 & 0.38 & 0.48 & 2.44 & 2.08 & 7.52 & 16.76 \\
5BZB J0006-0623 & 0.32 & 1.22 & 8.27 & 4.00 & 17.24 & 34.04 \\
5BZQ J1504+1029 & 0.31 & 0.66 & 9.44 & 2.52 & 10.36 & 25.24 \\
5BZU J0725-0054 & 0.31 & 0.62 & 14.03 & 4.00 & 7.78 & 19.74 \\
5BZQ J0108+0135 & 0.29 & 0.34 & 4.68 & 3.09 & 5.51 & 12.87 \\
5BZQ J1549+0237 & 0.28 & 0.26 & 1.23 & 4.00 & 5.59 & 13.04 \\
5BZQ J0501-0159 & 0.27 & 0.46 & 11.49 & 4.00 & 6.36 & 16.75 \\
5BZB J0831+0429 & 0.27 & 0.62 & 13.77 & 2.78 & 9.02 & 22.26 \\
5BZQ J0339-0146 & 0.26 & 0.60 & 11.39 & 4.00 & 8.24 & 19.69 \\
5BZB J1751+0939 & 0.23 & 0.33 & 2.74 & 2.65 & 7.35 & 16.65 \\
5BZQ J2301-0158 & 0.20 & 0.00 & 0.00 & 3.00 & 4.98 & 12.14 \\
5BZQ J0510+1800 & 0.20 & 0.00 & 0.00 & 3.25 & 6.69 & 18.84 \\
5BZB J0825+0309 & 0.19 & 0.78 & 19.82 & 3.30 & 9.61 & 23.76 \\
\textbf{5BZB J0509+0541} & \textbf{0.19} & \textbf{2.21} & \textbf{9.35} & \textbf{2.04} & \textbf{18.85} & \textbf{46.56} \\
5BZQ J0224+0659 & 0.18 & 0.21 & 0.32 & 3.25 & 6.82 & 15.07 \\
5BZQ J2148+0657 & 0.17 & 0.00 & 0.00 & 4.00 & 6.02 & 16.17 \\
5BZB J0238+1636 & 0.16 & 0.25 & 3.17 & 3.97 & 8.96 & 17.63 \\
5BZQ J2218-0335 & 0.15 & 0.47 & 3.56 & 4.00 & 7.12 & 15.26 \\
5BZQ J1224+0330 & 0.15 & 0.00 & 0.00 & 3.00 & 6.41 & 14.82 \\
5BZQ J2101+0341 & 0.14 & 0.00 & 0.00 & 4.00 & 5.42 & 12.92 \\
5BZQ J2229-0832 & 0.14 & 0.00 & 0.00 & 3.00 & 11.03 & 16.77 \\
5BZQ J1028-0236 & 0.14 & 0.00 & 0.00 & 4.00 & 5.10 & 11.40 \\
5BZQJ0217+0144 & 0.13 &  0.00 & 0.00 & 4.00 & 5.93 & 14.07 \\
5BZQ J2323-0317 & 0.13 & 0.00 & 0.00 & 3.25 & 5.43 & 12.51 \\
5BZB J2134-0153 & 0.12 & 0.42 & 8.55 & 2.78 & 6.11 & 15.08 \\
5BZQ J0532+0732 & 0.12 & 0.00 & 0.00 & 4.00 & 6.58 & 16.38 \\
\textbf{5BZU J1415+1320} & \textbf{0.12} & \textbf{2.00} & \textbf{20.76} & \textbf{2.44} & \textbf{20.64} & \textbf{50.60} \\
5BZQ J1224+2122 & 0.12 & 0.00 & 0.00 & 4.00 & 7.52 & 18.94 \\
5BZQ J2327+0940 & 0.12 & 0.00 & 0.00 & 4.00 & 7.03 & 16.79 \\
5BZB J0811+0146 & 0.12 & 1.48 & 31.86 & 3.97 & 13.39 & 33.64 \\
5BZB J0217+0837 & 0.11 & 0.00 & 0.00 & 4.00 & 6.68 & 15.91 \\
5BZQ J1222+0413 & 0.11 & 0.00 & 0.00 & 2.75 & 6.59 & 12.31 \\
5BZQ J0839+0104 & 0.11 & 0.00 & 0.00 & 3.75 & 5.36 & 13.57 \\
5BZQ J1833-2103 & 0.10 & 0.29 & 0.72 & 1.97 & 34.52 & 53.07 \\
5BZQ J0750+1231 & 0.10 & 0.68 & 15.23 & 3.05 & 11.68 & 28.64 \\
\hline \hline 
\end{tabular} 
\caption{ACT sources selected for baseline-subtracted individual searches. In this table we report best-fit results and 90$\%$ upper limits for sources tested with the baseline-subtracted temporal model. This model reflects the assumption that a source's neutrino flux increases in proportion with variability of the source beyond its quiescent state. Emission from extended regions beyond the blazar core are expected to contribute a steady-state mm flux. To correlate only with activity or particle injection at the variable jet base, the observed flux density minimum is taken to reflect this steady-state emission and subtracted. The two sources in excess of $2\sigma$ are bolded within the table. As described in Table~\ref{tab:upper_limit_lin}, the source name, ranking, local p-value, best fit signal parameters and upper limits are provided. Upper limits are provided in the form, $dN_{\nu}/dE_{\nu}(\mathrm{100 \,TeV}) = \phi_{90\%} \, \times$ $10^{-12}$\,$\mathrm{GeV}^{-1}$ $\mathrm{cm}^{-2}$, and represent the rate of neutrinos and antineutrinos of a single flavor.  }
\label{tab:upper_limit_baseline}
\end{table}

\begin{table}
\centering
\begin{tabular}{c c c c c c c}\hline \hline
Source Name &  \textrm{   } Model Rank (Index-Filtered) \textrm{   } & $-$log$_{10}(p_{\mathrm{local}}$) & \textrm{\,\,\,\,\,} $\hat{n}_{s}$ \textrm{\,\,\,\,\,} & \textrm{   } $\hat{\gamma}$ \textrm{   } & \textrm{\,\,\,\,} $\phi_{90\%, \gamma = 2.0}$ \textrm{\,\,\,\,} & \textrm{\,\,\,\,} $\phi_{90\%, \gamma = 2.5}$ \textrm{\,\,\,\,}   \\ \hline
5BZQ J1229+0203 & 1.00 & 0.98 & 1.63 & 1.93 & 6.84 & 14.42 \\
5BZU J1058+0133 & 0.48 & 0.00 & 0.00 & 4.00 & 2.72 & 6.76 \\
5BZQ J2232+1143 & 0.43 & 0.00 & 0.00 & 2.25 & 3.93 & 8.44 \\
5BZB J0006-0623 & 0.32 & 0.43 & 1.65 & 2.42 & 5.18 & 10.48 \\
5BZU J0725-0054 & 0.31 & 0.95 & 8.07 & 4.00 & 5.00 & 11.76 \\
5BZQ J1549+0237 & 0.29 & 0.00 & 0.00 & 4.00 & 3.09 & 6.36 \\
5BZQ J1743-0350 & 0.26 & 0.00 & 0.00 & 4.00 & 2.52 & 5.35 \\
5BZB J0854+2006 & 0.26 & 1.48 & 15.75 & 4.00 & 11.54 & 23.65 \\
5BZU J0433+0521 & 0.21 & 1.18 & 10.82 & 4.00 & 6.43 & 13.94 \\
5BZQ J0108+0135 & 0.20 & 0.00 & 0.00 & 3.25 & 2.90 & 6.48 \\
5BZQ J0501-0159 & 0.20 & 0.00 & 0.00 & 4.00 & 2.86 & 5.98 \\
5BZQ J1504+1029 & 0.20 & 0.47 & 2.40 & 3.49 & 4.95 & 10.33 \\
5BZQ J0423-0120 & 0.17 & 0.71 & 3.88 & 4.00 & 3.88 & 8.17 \\
5BZQ J0739+0137 & 0.16 & 0.00 & 0.00 & 4.00 & 2.73 & 6.33 \\
5BZB J1751+0939 & 0.16 & 1.29 & 6.55 & 2.63 & 8.39 & 19.42 \\
5BZQ J2301-0158 & 0.13 & 0.27 & 0.98 & 3.50 & 2.39 & 5.30 \\
5BZQ J0339-0146 & 0.13 & 0.76 & 4.36 & 4.00 & 4.52 & 9.22 \\
5BZQ J2148+0657 & 0.13 & 1.00 & 0.00 & 4.00 & 3.34 & 6.68 \\
5BZQ J2323-0317 & 0.12 & 0.00 & 0.00 & 2.50 & 2.79 & 6.29 \\
5BZQ J2136+0041 & 0.12 & 0.58 & 4.41 & 3.34 & 4.01 & 9.38 \\
5BZQ J2101+0341 & 0.12 & 1.18 & 11.74 & 4.00 & 6.82 & 16.51 \\
5BZQ J0224+0659 & 0.12 & 0.46 & 3.64 & 4.00 & 3.37 & 8.08 \\
5BZB J0831+0429 & 0.11 & 0.39 & 3.19 & 4.00 & 3.81 & 9.10 \\
5BZB J2134-0153 & 0.11 & 0.33 & 2.22 & 4.00 & 2.70 & 4.88 \\
5BZQ J0510+1800 & 0.11 & 0.65 & 4.92 & 2.99 & 6.03 & 11.63 \\
5BZQ J2218-0335 & 0.11 & 0.51 & 1.82 & 4.00 & 3.60 & 7.88 \\
5BZB J0825+0309 & 0.10 & 1.06 & 11.14 & 4.00 & 6.12 & 14.93 \\
\hline \hline 
\end{tabular} 
\caption{ACT sources selected for index-filtered individual searches. Here, results for sources analyzed with the index-filtered temporal model are reported. Best-fit results and 90$\%$ upper limits are provided. As TXS 0506+056 shows mm-spectral hardening during the period of the IC-170922A alert event, we consider hardened states for other sources within the catalog. Specifically, we choose to correlate with the top 32$\%$ of hardened activity. Emission following the 147\,GHz flux density is assumed only during this period, while no neutrino emission is assumed during other periods.  As described for Table~\ref{tab:upper_limit_lin}, the source name, ranking, local p-value, best fit signal parameters and upper limits are provided. Upper limit fluxes are presented in the form, $dN_{\nu}/dE_{\nu}(\mathrm{100 \,TeV}) = \phi_{90\%} \, \times$ $10^{-12}$\,$\mathrm{GeV}^{-1}$ $\mathrm{cm}^{-2}$, and represent the rate of neutrinos and antineutrinos of a single flavor. }
\label{tab:upper_limit_index}
\end{table}

\clearpage

\bibliographystyle{aasjournal}
\bibliography{neutrino}
\end{CJK*}
\end{document}